\documentclass[usenatbib,onecolumn]{mn2e}

\usepackage{graphicx}
\usepackage{color}
\usepackage{multirow}
\usepackage{amssymb}
\usepackage{amsmath}
\usepackage{anysize}
\usepackage{aas_macros}

\newcommand{\noplus}{}
\newcommand{\tmmathbf}[1]{\ensuremath{\boldsymbol{#1}}}

\renewcommand{\vec}[1]{\bmath{#1}}

\newcommand{\DS}{\displaystyle}

\renewcommand\Im{\operatorname{Im}}

\title[Instabilities in rotating jets]
      {Linear stability analysis of magnetized  jets: the rotating case}

\author[G.Bodo et al.] {G. Bodo$^{1}$\thanks{E-mail:
bodo@oato.inaf.it}, G. Mamatsashvili$^{2,3,4}$, P. Rossi$^{1}$ and A. Mignone$^{5}$\\
$^{1}$INAF/Osservatorio Astrofisico di Torino, Strada Osservatorio 20, 10025 Pino Torinese, Italy\\
$^{2}$Helmholtz-Zentrum Dresden-Rossendorf, P.O. Box 510119,
D-01314 Dresden, Germany\\
$^{3}$Department of Physics, Faculty of Exact and Natural Sciences, Tbilisi State University, \\
 Il. Chavchavadze ave. 3, Tbilisi 0179, Georgia\\
$^{4}$Abastumani Astrophysical Observatory, Ilia State University, G. Tsereteli str. 3, Tbilisi 0162, Georgia\\
$^{5}$Dipartimento di Fisica,  Universit\`a degli Studi di Torino, Via Pietro Giuria 1, 10125 Torino, Italy}

\begin{document}

\date{Accepted ??. Received ??; in original form ??}

\pagerange{\pageref{firstpage}--\pageref{lastpage}} \pubyear{2016}

\maketitle

\label{firstpage}

\begin{abstract}
We perform a linear stability analysis of magnetized rotating cylindrical jet flows in the approximation of zero thermal pressure. We focus our analysis on the effect of rotation on the current driven mode and on the unstable modes introduced by rotation. We find that rotation has a stabilizing effect on the current driven mode only for rotation velocities of the order of the Alfv\'en velocity.  Rotation introduces also a new unstable centrifugal buoyancy mode and the ``cold'' magnetorotational instability. The first mode is analogous to the Parker instability with the centrifugal force playing the role of effective gravity. The magnetorotational instability can be present, but only in a very limited region of the parameter space and is never dominant. The current driven mode is characterized by large wavelenghts and is dominant at small values of the rotational velocity, while the buoyancy mode becomes dominant as rotation is increased and is characterized by small wavelenghts.   
\end{abstract}


\section{Introduction}
%
%
%

An important step for understanding the dynamics and phenomenology of astrophysical jets is the study of their instabilities. Instabilities have a substantial importance, on one hand for the formation and evolution of various observed structures and, on the other hand, for dissipating part of the jet energy and leading to the observed radiation. There are several possible sources of instabilities, like the velocity shear between the jet and the ambient medium, which drives the Kelvin-Helmholtz instability, the current flowing along magnetic field lines, which drives the current driven instability (CDI) and rotation that can drive several kinds of instabilities. Since the most promising models for the acceleration and collimation of jets involve the presence of a magnetic field with footpoints anchored to a rotating object (an accretion disk or a spinning star or black hole), the presence of a toroidal field component and of rotation seems to be a natural consequence and both CDI and rotation driven instabilities may play an important role in the jet propagation. CDI have been, for example, suggested as being responsible  for  the conversion from  Poynting  to kinetic energy flux  in the first phases of jet propagation \citep{Sikora05}.

 KHI have been extensively studied in several different configurations both in the Newtonian \citep[see e.g.][]{Bodo89, Birkinshaw91, Hardee92, Bodo96, Hardee06, Kim15} and relativistic \citep[see e.g.][]{Ferrari78, Hardee79,  Urpin02, Perucho04, Perucho10, Hardee07} cases. Similarly, CDI  have been widely studied in the Newtonian limit \citep[see e.g.][]{Appl92, Appl96, Begelman98, Appl00, Baty02, Bonanno11, Bonanno11a},  while the analysis of the relativistic case has been more limited,  most of the studies have considered the force-free condition \citep{Pariev94, Pariev96, Lyubarski99, Tomimatsu01, Narayan09} and only \citet{Bodo13} studied the full MHD case. The study of the effects of rotation have been mainly focused on the accretion disk problems, where the main instability considered is the magnetorotational instability \citep{Balbus92}, however, the combination of magnetic field and rotation can give rise to several other instabilities   \citep[see e.g.][]{Kim00,  Hanasz00, Keppens02, Varniere02, Huang03,Pessah05, Bonanno06, Bonanno07, Fu11} and the interplay between the different modes can become quite complex. Our goal is to study these rotation-induced instabilities in the context of magnetized jets.

In \citet{Bodo13} (hereinafter Paper I)  we studied the interplay between KHI and CDI in a relativistic non-rotating cold jet configuration, characterized by a current distribution concentrated inside the jet and closing at large distances. In this paper, we introduce the effects of rotation which, however, makes the analysis of the unstable modes much more intricate. Therefore,  before tackling the full relativistic case, in this paper we limit ourselves to a newtonian analysis, neglecting again thermal pressure compared with the magnetic one. Moreover, since most of the unstable modes that we will consider are concentrated  inside the jet radius and therefore the effect of the jet velocity would be to simply Doppler shift their frequencies, in this first step we ignored also the presence of the longitudinal velocity component. The main focus of this paper will then be on the effect of rotation on CDI and on the new modes of instability introduced by rotation. The effect of rotation on CDI was considered by \citet{Carey09} who analyzed a rigidly rotating jet and found a stabilizing effect for rotation periods shorter than a few Alfv\'en times. An analysis of the unstable  modes introduced by rotation in a configuration and parameter range similar to ours has been performed by \citet{Kim00}. They  discuss these modes in the cold plasma limit, however, their study is mainly local, whereas we focus more on global analysis of these instabilities, besides they do not discuss the CDI. Another related works are by \citet{Pessah05} and \citet{Huang03}, who examined the instabilities of axisymmetric perturbations in the presence of rotation and superthermal magnetic fields. The treatment of the first paper is again local and mostly focuses on an equilibrium configuration typical of  accretion discs, while the second one analyses the stability a rotating cylindrical plasma Dean flow with only axial field both with local and global approach.      

The plan of the paper is the following: in the next section we present the equilibrium configuration, in Section \ref{sec:linequations} we derive the linearized equations,  in Section \ref{modeclassification} we discuss the WKBJ local dispersion relation and present energetic considerations based on the Frieman-Rotenberg approach \citep{Frieman60} . The local dispersion relation and the energetic considerations will be useful in understanding the nature of the unstable modes that will be discussed in Section \ref{results}, where we present our results on global modes. Finally in the last Section \ref{summary} we summarize our findings.

\section{Problem Description}
\label{problem}
%
%
%

We study the linear stability of a cold magnetized cylindrical jet flow. Although in the following we will consider only the zero thermal pressure case, we keep here the presentation more general. The relevant equations are  the equations of ideal MHD:
\begin{equation}\label{eq:drho/dt}
\frac{\partial  \rho }{\partial t}  + \nabla \cdot ( \rho  \vec{v}) = 0 \,,
\end{equation}
\begin{equation}\label{eq:dm/dt}
 \rho \frac{\partial  \vec{v} }{\partial t} +  \rho (\vec{v} \cdot \nabla )  \vec{v} = - \nabla\left(p + \frac{B^2}{2}\right) + (\vec{B} \cdot \nabla) \vec{B} \,
\end{equation}
\begin{equation}\label{eq:dB/dt}
\frac{\partial   \vec{B}}{\partial t} =  \nabla \times (\vec{v} \times \vec{B})   \,,
\end{equation}
\begin{equation}\label{eq:dp/dt}
\frac{\partial p}{\partial t} + \vec{v}\cdot\nabla p + c_s^2\rho \nabla\cdot\vec{v} = 0 \,,
\end{equation}
where $\rho$ is the  density,  $p$ is the pressure, $c_s$ is the sound speed,  $\vec {v}$, $\vec {B}$, are, respectively, the velocity and magnetic fields.  We  remark that a factor of $\sqrt{4 \pi}$ is absorbed in the definition of  $\vec{B}$. The first step in the stability analysis is to define an equilibrium state satisfying the stationary form of equations (\ref{eq:drho/dt}-\ref{eq:dp/dt}) and this will be done in the next subsection.

\subsection{Equilibrium Configuration}
%
%
\label{sec:equilibrium}

We adopt a cylindrical system of coordinates $(r,\varphi,z)$ (with versors $\vec{e_r}, \; \vec{e_\varphi}, \; \vec{e_z}$)
and seek for  axisymmetric steady-state solutions,
i.e., $\partial_t = \partial_\varphi = \partial_z=0$.  The jet propagates in the vertical ($z$) direction, the magnetic field  and velocity have no radial component and consist of a vertical (poloidal) component  $B_z, v_z$, and a toroidal component $B_\varphi, v_\varphi$. The magnetic field configuration can be characterized by the pitch parameter
\begin{equation}\label{eq:pitch}
  \hat P = \frac{r B_z}{B_\varphi} \,.
\end{equation}

The only non-trivial equation is given by the radial component of the momentum equation (\ref{eq:dm/dt}) which, in the zero pressure case, simplifies to
\begin{equation}\label{eq:radial_eqMHD}
  \rho v_\varphi^2 = \frac{1}{2r}\frac{d(r^2B_\varphi^2)}{dr} 
  + \frac{r}{2}\frac{dB_z^2}{dr} \,.
\end{equation}
Equation (\ref{eq:radial_eqMHD}) leaves the freedom of choosing the radial profiles of all flow variables but one and then solve for the remaining profile. Furthermore, we note that the presence of a longitudinal velocity has no effect on the radial equilibrium.  

The choice of the radial profiles is somewhat arbitrary since we have no direct information about the magnetic configuration in astrophysical jets. The choice of the $B_\varphi$ distribution is equivalent to a choice of the distribution of the longitudinal component of the current and also determines the behavior of the pitch parameter $P(r) $, that is important for the stability properties.  In principle, one can then have several equilibria characterized by different forms of the current distribution, that can be  more or less concentrated, can peak on the axis or at the jet boundary,  and  can close in different ways \citep[see e.g.][]{Appl00, Bonanno08, Bonanno11, Kim15}.  Our choice is to consider a general class of constant density equilibria in which the vertical current density is peaked on the axis and is concentrated in a region of radius $a$.  The azimuthal component of magnetic field has therefore to behave linearly with radius close to the origin and decay as $1/r$ at large distances, more precisely we can write it as
\begin{equation}
  B^2_{\varphi} = \frac{H^2_c}{a^2} \frac{a^2}{r^2} f \left( \frac{r}{a} \right)
   \label{eq:bphi}
\end{equation}
where the function $f$ behaves in the following way at small and large radii:
\begin{equation}
f \approx \left( \frac{r}{a} \right)^4  \qquad \hbox{for} \quad r \rightarrow 0;   \qquad\qquad f \rightarrow 1 \qquad \hbox{for} \quad r \rightarrow \infty.
\end{equation}
 
 For the rotational frequency  $\omega = v_\varphi / r$ we assume the form
\begin{equation}
  \Omega^2 = \frac{\hat \Omega^2_c}{4}  \left( \frac{a}{r} \right)^3 \frac{d f (r/a)}{d (r/a)},
   \label{eq:omega}
\end{equation}
where $\hat \Omega_c$ is the value of $\Omega$ on the axis, i.e. $\hat \Omega_c=\Omega(0)$. We make this assumption for simplicity and for avoiding any possible non-monotonic behaviors of  $B_z$. 
 
Inserting expressions (\ref{eq:bphi}) and (\ref{eq:omega}) in the equilibrium condition (\ref{eq:radial_eqMHD}), we can get the profile of $B_z$ as  
\begin{equation}\label{eq:bz}
 B^2_z = \frac{\hat P_c H^2_c}{a^4} -  \frac{H^2_c}{a^2}  (1 - \alpha ) F \left( \frac{r}{a} \right) 
\end{equation}   
where
\begin{equation}
\hat P^2_c = \lim_{r \rightarrow 0}  \frac{r^2 B^2_z}{B^2_{\varphi}},
\end{equation}
\begin{equation}\label{eq:alfa}
\alpha = \frac{\rho  {\hat \Omega^2_c} a^4}{2 H_c^2}
\end{equation}
and
\begin{equation}
F\left(\frac{r}{a}\right) =  \int_0^{r/a} \frac{1}{\xi^2} \frac{d f(\xi)}{d \xi} d \xi
\end{equation}
is a monotonic function with $F(0) = 0$ and $ \lim_{r \rightarrow \infty} F = F_\infty$. Depending on the value of $\alpha$, $B_z$ is  decreasing (for $\alpha < 1$) or increasing (for $\alpha > 1$). From a physical point of view, if we look at equation (\ref{eq:radial_eqMHD}), we can see that the equilibrium is given by the balance of three forces: gradient of  $r^2 B^2_\varphi$, the gradient  of $B^2_z$ and  the centrifugal force. When there is no rotation, the gradient  of  $r^2 B^2_\varphi$, which always points towards the jet axis, is balanced by the gradient of  $B^2_z$.  Increasing the rotation rate, the gradient of  $B^2_z$ decreases, until, for $\alpha = 1$, $B_z$ becomes constant. If we still increase rotation beyond this point, the centrifugal term becomes larger than the gradient of  $r^2 B^2_\varphi$, thus $B_z$ has to increase outward for providing an inward force term, needed for having equilibrium. 

We can now define a radially averaged Alfv\'en velocity via
\begin{equation}\label{eq:Bav}
 \langle v_A\rangle^2 \equiv \frac{\int_0^{a} (B_z^2 + B_\varphi^2)r\,dr}{\rho \int_0^{a} r \,dr}.
\end{equation}
and inserting  the expressions for $B_\varphi$ and $B_z$ from equations  (\ref{eq:bphi}) and (\ref{eq:bz}) in equation (\ref{eq:Bav}), we get
\begin{equation}
\frac{2H^2_c}{\rho a^2  \langle   v_A\rangle^2} I_1 \noplus + \frac{\hat P^2_c H^2_c}{\rho a^4  \langle  v_A \rangle^2} + \frac{ {\hat \Omega}^2_c  a^2}{ \langle  v_A\rangle^2} I_2 - \frac{2H^2_c}{ \rho  a^2 \langle  v_A\rangle^2} I_2  = 1 
\end{equation}
where 
\[ 
I_1 = \int^1_0 \frac{1}{\xi} f \left( \xi \right) d \xi  \hspace{2em} 
I_2 = \int^1_0 \xi F \left( \xi \right)   d \xi 
\]
from which we get
\begin{equation}\label{eq:Hc}
 \frac{H^2_c}{\rho a^2  \langle  v_A\rangle^2} = \frac{1- I_2 \Omega^2_c}{P^2_c + 2 I_1 - 2 I_2}
\end{equation}
where 
\begin{equation}\label{eq:param}
\Omega_c = \frac{ {\hat \Omega_c} a}{\langle  v_A \rangle},   \hspace{2em} P_c = \frac{\hat P_c}{a}
\end{equation}
are, respectively, nondimensional measures of the rotation rate and of  the value of the pitch on the axis. For a given choice of the  function $f$ and the values of these two parameters, we can derive the value of $H_c$ from equation (\ref{eq:Hc}) and the equilibrium structure is then fully determined. However not all combinations of $\Omega_c$ and $P_c$ are allowed, because, in order to have a physically meaningful solution, we have to impose the additional constraints that $B_\varphi^2$ and $B_z^2$ have to be everywhere positive, which translate as  
\begin{equation}
H_c^2>0,~~~2P_c^2 H^2_c+  \left( \rho {\hat\Omega}^2_c a^4- 2 H^2_c \right) F_\infty > 0,
\end{equation}
respectively, for equations (\ref{eq:Hc}) and (\ref{eq:bz}). 

In order to exemplify the structure of the equilibrium solution, we now make a specific choice for the function $f$, 
\begin{equation}
f\left( \frac{r}{a} \right) = 1 - \exp\left(-\frac{r^4}{a^4}\right)
\end{equation}
which gives the same solution used in Paper I. This same solution will also be used in the next sections for computing the instability behavior. More specifically, we have the following profiles for $B_\varphi^2$, $B_z^2$ and $\Omega^2$
\begin{equation}
B_\varphi^2 = \frac{H^2_c}{a^2} \frac{a^2}{r^2} \left[1 - \exp\left(-\frac{r^4}{a^4}\right) \right],
\end{equation}
\begin{equation}
B_z^2 =  \frac{\hat P_c H^2_c}{a^4}  - (1 - \alpha) \frac{H_c^2\sqrt{\pi}}{a^2}{\rm erf}  \left(\frac{r^2}{a^2}\right)
\end{equation}
where $\mathrm{erf}$ is the error function, and
\begin{equation}
\Omega^2 =  {\hat \Omega^2_c} \exp\left(-\frac{r^4}{a^4}\right)
\end{equation}
\begin{figure*}
 \centering
 \includegraphics[width=15cm]{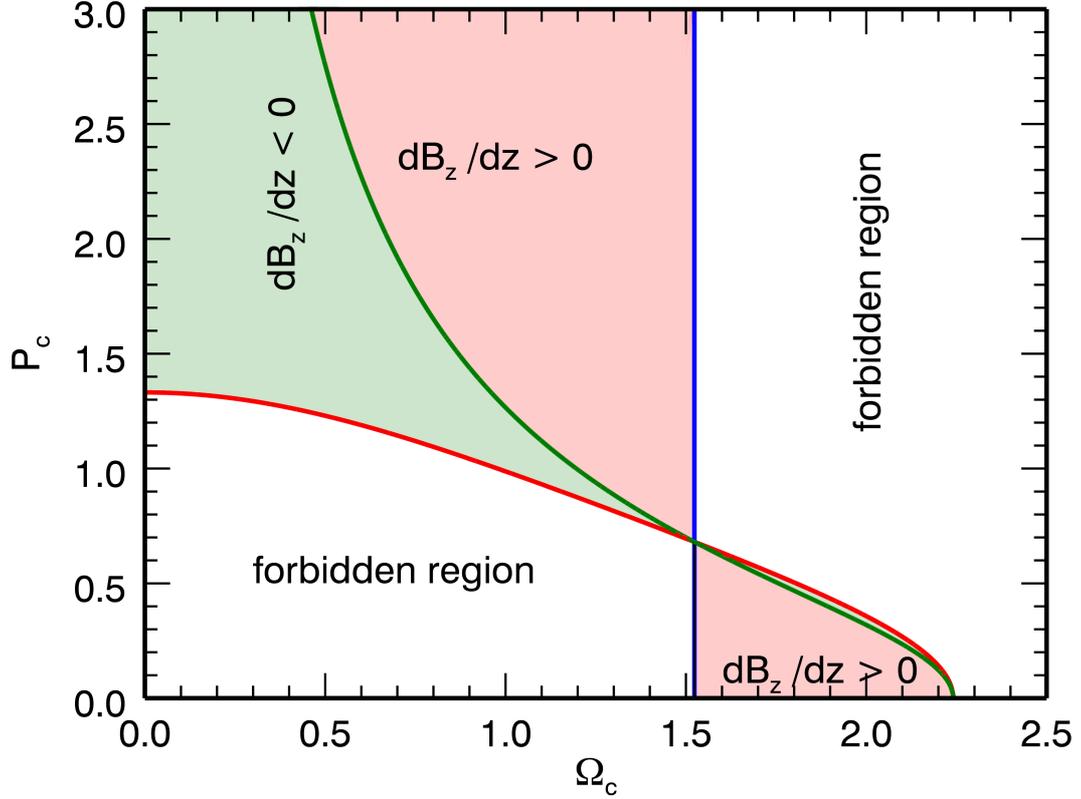} 
 \caption{\small  Regions (colour shading) of allowed equilibrium solutions in the parameter plane $(\Omega_c, P_c)$. The red and blue curves mark the boundaries of this region and represent, respectively, the conditions $\lim_{r \rightarrow \infty} B_z = 0$ and $H_c=0$. The green curve represents the points where $\alpha = 1$ and $B_z$ is constant with radius. We show by a green shading the region where $d B_z / d r < 0$ and by a red shading the region where    $d B_z / d r  >  0$.}
 \label{fig:equil}
\end{figure*}
In Fig. 1, we present the regions in the $(\Omega_c , P_c)$ plane for which the equilibrium is possible. The red curve represents the condition $B_z^2  = 0$, while the blue curve represents the condition $H_c^2 > 0$ and the green curve represents the combinations of $\Omega_c$ and $P_c$ for which $B_z$ is constant. Therefore the green region represents equilibria for which $dB_z / dz < 0$, while in the red region  $dB_z / dz >0$.

We conclude by summarizing the parameters determining the equilibrium: once the magnetization radius $a$ and the average Alfv\'en velocity, $\langle v_A \rangle$, are fixed, the equilibrium structure is fully determined by the two parameters $P_c$ and $\Omega_c$. We also observe that an arbitrary profile of the longitudinal velocity $v_z$ can be superposed to the equilibrium, however in this paper we consider only the case $v_z = 0$.

\section{Linearized Equations}
\label{sec:linequations}
We consider small perturbations of the form  $\propto \exp \left( {\rm i} \omega t -{\rm i} m \varphi -{\rm i} k z \right)$ to the equilibrium state described above. By linearizing the ideal MHD equations, we obtain the following system of two first order ordinary differential equations 
\begin{equation}\label{eq:lin_system1}
\Delta\frac{d\xi_{1r}}{dr}=\left(C_1-\frac{\Delta}{r}
\right)\xi_{1r}-C_2P_1,
\end{equation}
\begin{equation}\label{eq:lin_system2}
\Delta\frac{dP_1}{dr}=C_3\xi_{1r}-C_1P_1,
\end{equation} 
where $\xi_{1r}$ is the radial component of the Lagrangian displacement related to the Eulerian perturbation of
the velocity field $\tmmathbf{v}_1$ through
\begin{equation}\label{eq:displ}
\tmmathbf{v}_1=\left(\frac{\partial}{\partial t}+\tmmathbf{v}_0\cdot
\nabla\right)\tmmathbf{\xi}_1-(\tmmathbf{\xi}_1\cdot\nabla)\tmmathbf{v}_0,
\end{equation}
and $P_1$ is the  total pressure perturbation
 \begin{equation}
P_1=c_s^2\rho_1+\tmmathbf{B}_0\cdot\tmmathbf{B}_1,
\end{equation}
with $\rho_1$ and $\tmmathbf{B}_1$ being respectively the density and magnetic field perturbations.  In equations (\ref{eq:lin_system1}) and (\ref{eq:lin_system2}),  $\Delta$, $C_1$, $C_2$ and $C_3$ depend on the equilibrium quantities and on $\omega$, $k$, $m$ and are defined as
\begin{equation}
\begin{split}
\Delta=\rho_0(B_0^2+\rho_0c_s^2)\tilde{\omega}^4-k_B^2(B_0^2+2\rho_0c_s^2)\tilde{\omega}^2+c_s^2k_B^4= 
[\rho_0\tilde{\omega}^2-k_B^2][\tilde{\omega}^2(B_0^2+\rho_0c_s^2)-c_s^2k_B^2],
\end{split}
\end{equation}
\begin{equation}
\begin{split}
C_1=\frac{\rho_0\tilde{\omega}^2}{r}[\tilde{\omega}^2(B_{0\varphi}^2-\rho_0v_{0\varphi}^2)+(\tilde{\omega}B_{0\varphi}+v_{0\varphi}k_B)^2] 
- \frac{2m}{r^2}(k_BB_{0\varphi}+\rho_0v_{0\varphi}\tilde{\omega})[\tilde{\omega}^2(B_0^2+\rho_0c_s^2)-c_s^2k_B^2],
\end{split}
\end{equation}
\begin{equation}
C_2=\rho_0\tilde{\omega}^4-\left(k^2+\frac{m^2}{r^2}\right)[\tilde{\omega}^2(B_0^2+\rho_0c_s^2)-c_s^2k_B^2],
\end{equation}


\begin{equation}
  \begin{split}
C_3  = \DS    \Delta\left[\rho_0\tilde{\omega}^2-k_B^2+r\frac{d}{dr}   \left(\frac{B_{0\varphi}^2-\rho_0v_{0\varphi}^2}{r^2}\right) \right] 
      -\frac{4[\tilde{\omega}^2(\rho_0c_s^2+B_0^2)-c_s^2k_B^2](k_BB_{0\varphi}
      +\rho_0v_{0\varphi}\tilde{\omega})^2}{r^2} \\
      +\frac{\rho_0[\tilde{\omega}^2(B_{0\varphi}^2-\rho_0v_{0\varphi}^2)
           +(\tilde{\omega}B_{0\varphi}+k_Bv_{0\varphi})^2]^2}{r^2},
  \end{split}
\end{equation}
where quantities with $0$ subscript refer to the equilibrium state and
\begin{equation}
\tilde{\omega}\equiv\omega-\frac{m}{r}v_{0\varphi}-kv_{0z},~~~k_B\equiv\frac{m}{r}B_{0\varphi}+kB_{0z}.
\end{equation}
We note that the system (\ref{eq:lin_system1}) and (\ref{eq:lin_system2}) was derived by \citet{Bondeson87} and we kept it in its general form even though in the following we will consider only the case with $v_{0z} = 0$ and $c_s = 0$. This system, supplemented with appropriate boundary conditions at $r=0$ and $r \rightarrow \infty$, poses an eigenvalue problem for $\omega$.  On the axis, at $r=0$, the equations are singular but the solutions have to be regular, while at infinity the solutions have to decay and no incoming wave is allowed (Sommerfeld condition). This asymptotic  behaviour of the solutions for small and large radii are used in the numerical integration of the eigenvalue problem. For finding eigenvalues we use a shooting method with a complex secant root finder, as we did in Paper I. The numerical integration cannot start at $r=0$ because of the singularity, so we start at a small distance from the origin where the solution is obtained through a series expansion of the equations described in the  Appendix \ref{ap:small_r}. Similarly, we start a backward integration from a sufficiently large radius, where the asymptotic solution is obtained as described in the Appendix \ref{ap:large_r} and then we match the two numerical solutions at an intermediate radius. 

Equations (\ref{eq:lin_system1}) and (\ref{eq:lin_system2})  have singularities whenever $\Delta = 0$ which give rise to four distinct continua, two Alfv\'en continua for
\begin{equation}
\omega = \frac{m}{r}v_{0\varphi} + kv_{0z} \pm \frac{k_B}{\rho_0}
\end{equation}
and two slow continua for 
\begin{equation}
\omega = \frac{m}{r}v_{0\varphi} + kv_{0z} \pm \frac{k_B^2 c_s^2}{B_0^2 + \rho_0 c_s^2}
\end{equation}
In the case of zero pressure and zero longitudinal velocity, that we consider in this paper, the slow continua reduce to the single flow continuum defined by the condition
\begin{equation}
\omega = \frac{m}{r}v_{0\varphi} 
\end{equation}
Since we are interested only in unstable solutions, our integration path will always avoid the singularities, however the presence and position of the continua is fundamental in shaping the overall MHD spectrum \citep{Goedbloed10}.

\section{Classification of modes}
\label{modeclassification}
%
%
%

To classify the unstable modes present in the jet and understand their physical origin, following \cite{Kim00, Keppens02, Blokland05,Pessah05,Goedbloed09}, we employ the WKBJ approach and energetic considerations following  the Frieman-Rotenberg formalism \citep{Frieman60}. The combination of these methods allows us to gain insight into the nature of dominant driving forces inducing the instability of each mode and classify them accordingly.

\subsection{WKBJ approach}
\label{sec:wkb}

Equations (\ref{eq:lin_system1}) and (\ref{eq:lin_system2}) can be combined in a single second-order differential equation only for ${\xi_{1r}}$,
\begin{equation}\label{eq:secondorder}
\begin{split}
\frac{d^2}{dr^2}(r\xi_{1r})+\frac{d}{dr}\ln\left(\frac{\Delta}{rC_2}\right)\frac{d}{dr}(r\xi_{1r}) 
+\left[\frac{C_2C_3-C_1^2}{\Delta^2}-\frac{rC_2}{\Delta}\frac{d}{dr}\left(\frac{C_1}{rC_2}\right)\right](r\xi_{1r})=0.
\end{split}
\end{equation}
Assuming the radial wavelength of perturbations  small compared to the length scale over which there are significant variations in the equilibrium quantities, we can represent the radial dependence of the displacement as $\xi_{1r}\propto \exp({\rm i}\int k_r(r')dr')$, where the radial wavenumber $k_r$ is assumed to be large, $rk_r \gg 1$. Substituting this into equation (\ref{eq:secondorder}) and neglecting the radial variations of the equilibrium quantities,  we obtain, to leading order in the large parameter $rk_r$, the following local dispersion relation \citep[see e.g.,][]{Keppens02,Blokland05,Pessah05}
\begin{equation}
k_r^2\Delta^2+C_1^2-C_2C_3=0,
\end{equation}
which after using the expressions for $C_1, C_2, C_3$ and $\Delta$ reduces to a sixth-degree polynomial 
\begin{equation}\label{eq:fullwkb}
\begin{split}
\rho_0^2\tilde{\omega}^6    -        \rho_0\tilde{\omega}^4\left[k_B^2+4\rho_0\Omega^2+r\frac{d}{dr}(\rho_0\Omega^2) + k_t^2 B_0^2 \right]  
 - 4\rho_0\Omega \left( \frac{m}{r} \rho_0(r\Omega^2) + 2\Omega_B kB_{0z} \right)  \tilde{\omega}^3   \\ 
+  \tilde{\omega}^2 \bigg[   k_t^2 k_B^2B_0^2 +\left(k^2+\frac{m^2}{r^2}\right)    \Big( 4\rho_0\Omega^2B_{0\varphi}^2  -\rho_0^2(r\Omega^2)^2    
+ B_0^2r \frac{d}{dr} \left( \rho_0\Omega^2 -\Omega_B^2 \right) +4 \rho_0 B_{0\varphi} \Omega_B (r\Omega^2)  \Big)  \\   
- 4\frac{m}{r}  \rho_0 k_B \Omega_B  (r\Omega^2)  - 4 B_0^2  k^2\Omega_B^2   \bigg]  
+4\rho_0B_{0\varphi}k_B\Omega(r\Omega^2)\left(k^2+\frac{m^2}{r^2}\right)\tilde{\omega}  
+\rho_0(r\Omega^2)^2k_B^2\left(k^2+\frac{m^2}{r^2}  \right)=0,
\end{split}
\end{equation}
where $\Omega_B=B_{0\varphi}/r$ describes the effect of the curvature of  toroidal field lines in cylindrical geometry and $k_t^2 = k_r^2 + k^2 +m^2/r^2$ is the total wavenumber squared. We intentionally separated out the product $\rho_0(r\Omega^2)$, which represents the centrifugal force per unit volume. 

This general dispersion relation describes, in principle, all the modes in the local approximation, however, the coefficients are complicated and not physically revealing. So, we examine various limiting cases to identify these modes. We start by considering the limits for large and small wavenumbers.

At large $k$, when $kr\gg|m|$ and $k^2 v_A^2\gg \rho_0\tilde{\omega}^2$, to leading order from (\ref{eq:fullwkb}) we get
\begin{equation}\label{eq:pbm}
\tilde{\omega}^2=-\frac{(r\Omega^2)^2}{v_A^2}\frac{k^2}{k_r^2+k^2},
\end{equation}
where $v_A$ is the local Alfv\'en velocity. This dispersion relation resembles that of the Parker instability with the driving role of external gravity replaced here by the centrifugal force per unit mass $(r\Omega^2)$ \cite[see e.g.,][]{Huang03}. Thus, the unstable mode at large $k$, is driven mainly by the centrifugal force and can be identified with a magnetic buoyancy mode \citep{Kim00}, which in this limit of large vertical wavenumber operates by bending mostly poloidal field lines. The growth rate, $\gamma=-\Im(\omega)$, for $k\rightarrow \infty$ tends asymptoticallly to its maximum value $\gamma_{max}=r\Omega^2/v_A$.

At small $k\ll |m|/r$ and for $k_B^2 \gg \rho_0\tilde{\omega}^2$, from equation (\ref{eq:fullwkb}) we get the following dispersion relation
\begin{equation}\label{eq:tbm}
{\cal A} \tilde{\omega}^2+4\Omega(r\Omega^2)\tilde{\omega}\frac{m}{r}+(r\Omega^2)^2\frac{m^2}{r^2}=0,
\end{equation}
where
\begin{equation}
{\cal A} = \left[v_A^2\left(k_r^2+\frac{m^2}{r^2}\right)+\kappa^2+\frac{B_{0z}^2}{B_{0\varphi}^2} r\frac{d}{dr}(\Omega^2)\right]
\end{equation}
and $\kappa$ is the epicyclic frequency, $\kappa^2=4\Omega^2+2r\Omega d\Omega/dr$.
The solution is given by 
\begin{equation}\label{eq:tbmsol}
\tilde{\omega} = \frac{-2 m r\Omega^3 \pm m\sqrt{4 r^2 \Omega^6 - {\cal A} (r \Omega^2)^2}}{r{\cal A}}
\end{equation}
It is clear that if ${\cal A}$ is positive, which, as we will show below, corresponds to the jet flow being  stable against the cold magnetorotational instability (MRI), only the last term of equation (\ref{eq:tbm}),  proportional to the square of the centrifugal acceleration, guarantees the existence of instability. As a result, also in this limit, we have again the centrifugal buoyancy mode, but now, at small $k$, which mainly operates by bending toroidal field lines. In fact, the dispersion relation (\ref{eq:tbm}), in the $B_{0\varphi}$-dominated regime, is similar to that of the non-axisymmetric toroidal buoyancy mode derived in \cite{Kim00} (see their equation 51). Note that this approximation can work only for non-axsymmetric modes, since it is based on the condition $k \ll |m|/r$, that cannot be satisfied for $m=0$. On the other hand, in the axisymmetric case, it is not possible to bend toroidal field lines. The instability condition for this mode can be derived from equation (\ref {eq:tbmsol}) as
\[
4\Omega^2-\kappa^2-\frac{B_{0z}^2}{B_{0\varphi}^2}r\frac{d\Omega^2}{dr}<v_A^2\left(k_r^2+\frac{m^2}{r^2}\right)
\]

\begin{figure*}
   \centering
   \includegraphics[width=15cm]{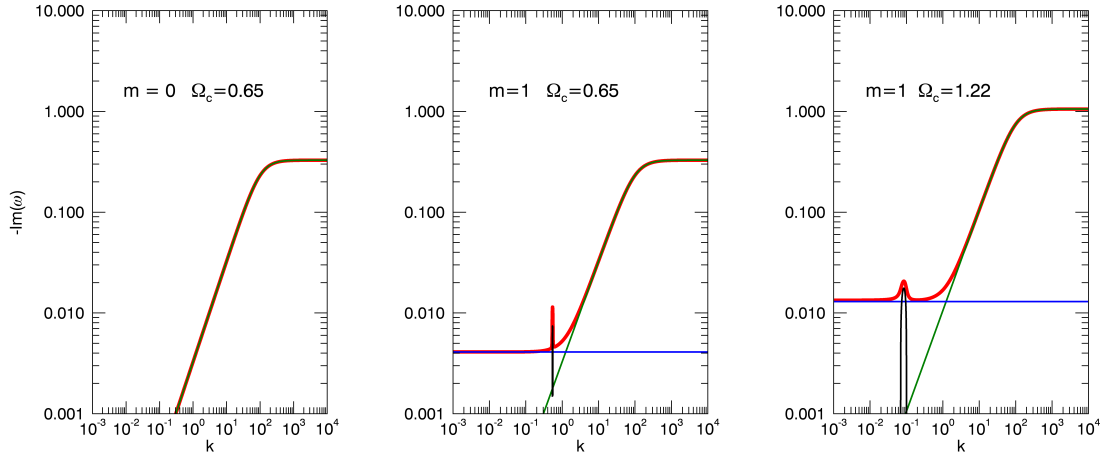} 
   \caption{\small Plot of the growth rate of unstable modes as a function of the wavenumber in the WKBJ approximation. The red lines represent the numerical solution to equation (\ref{eq:fullwkb}), the green curve represents the analytical approximation to the poloidal buoyancy mode given by equation (\ref{eq:pbm}), the blue line represents the analytical approximation to the toroidal buoyancy mode given by equation (\ref{eq:tbm}) and, finally, the black curve  represents the analytical approximation to the magneto-rotational mode given by equation (\ref{eq:mri}). The parameters for the solutions are $P_c = 1.66$, $k_r=100$ and the solutions are computed for the radial position $r=0.8$. The left panel is for $m=0$ and a rotation rate $\Omega_c = 0.65$ . The mid and right panels are for $m=1$  an two different values for $\Omega_c$,   $\Omega_c = 0.65$ for the mid panel and $\Omega_c = 1.22$ for the right panel. }
   \label{fig:figwkbj}
\end{figure*}

In principle, the cold differentially rotating jet can also support the MRI arising from the combined effect of differential rotation and magnetic fields. To capture this instability, in equation (\ref{eq:fullwkb}) we ignore centrifugal $r\Omega^2$ and curvature $\Omega_B$ terms, which are not its main driving factors, but retain rotation $\Omega$ (i.e., Coriolis force) and shear $d\Omega/dr$, which together with azimuthal and vertical magnetic fields cause this instability. As a result, we obtain a more compact dispersion relation describing the MRI in cold differentially rotating cylindrical flows \cite[see also][]{Kim00,Pessah05}
\begin{equation}\label{eq:mri}
\begin{split}
\rho_0^2 \tilde{\omega}^4 -  \rho_0\tilde{\omega}^2   \left[   k_B^2 + 4\rho_0\Omega^2 + r  \frac{d}{dr}(\rho_0\Omega^2)  + \left( k^2+k_r^2+\frac{m^2}{r^2} \right) B_0^2 \right]  \\
 + \left(k_r^2+k^2+\frac{m^2}{r^2}\right)  k_B^2B_0^2  +  \left(k^2+\frac{m^2}{r^2}\right)   \left(4 \rho_0\ Omega^2 B_{0\varphi}^2 + B_0^2 r \frac{d}{dr}   ( \rho_0 \Omega^2 )  \right) = 0
\end{split}
\end{equation}
This expression is a quadratic polynomial for $\tilde{\omega}^2$ from which a condition for the cold MRI can be readily deduced. That is, an unstable solution $\tilde{\omega}^2<0$ can exist whenever
\[
\frac{k_r^2k_B^2}{k^2+m^2/r^2}+k_B^2+\rho_0\kappa^2<4\rho_0\Omega^2\frac{B_{0z}^2}{B_0^2}.
\] 
The most favourable condition for the instability is when $k_B = 0$ and from this we can derive the necessary condition $\kappa^2B_0^2< 4\Omega^2B_{0z}^2$. Hence, in the cold plasma limit ($c_s=0$), the MRI vanishes in the case of a purely toroidal field and the presence of a nonzero poloidal/vertical field component is necessary for its operation.

In Fig. \ref{fig:figwkbj} we show a comparison between representative  full numerical solutions to equation (\ref{eq:fullwkb}),  and the analytical approximations given by equations (\ref{eq:pbm}), (\ref{eq:tbmsol}) and (\ref{eq:mri}). The left panel is for the axisymmetric mode with $P_c = 1.66$ and $\Omega_c = 0.65$. In this case we have only the poloidal buoyancy mode, the numerical solution is represented by the red curve, while the analytical approximation, given by equation (\ref{eq:pbm}), is represented by the green curve and the two curves are indistinguishable. The mid and right panels are for $m=1$, the same value of $P_c$ and two different values of the rotation rate, $\Omega_c=0.65$ for the mid panel and $\Omega_c=1.22$ for the right panel. Again, the numerical solution is represented by the red curve and, in this case, we have the analytical approximations for the poloidal buoyancy mode (equation \ref{eq:pbm}), at large wavenumbers, represented by the green curve, and for the toroidal buoyancy mode (equation \ref{eq:tbm}) at small wavenumbers, represented by the blue curve. We can furthermore notice that around the value of $k$ for which $k_B = 0$, we have a narrow peak in the growth rate, which corresponds to a very localized MRI. The peak is very narrow for $\Omega_c = 0.65$ and widens with increasing the rotation rate (right panel with $\Omega_c=1.22)$. The black curve, representing the solution to equation (\ref{eq:mri}), reproduces the behavior of the numerical solution, although the agreement is not as good as for the other two approximations due to the terms neglected in the derivation of equation (\ref{eq:mri}).The solutions are taken at a particular radial position, $r=0.8$, however, taking different radial positions, the qualitative behavior of the solution remains the same. 

\subsection{Energetic considerations}

Alternatively, the classification of modes and related instabilities performed above using the WKBJ approach can also be made based on energetic considerations that can be derived following the Frieman-Rotenberg formalism  \citep{Frieman60, Freidberg87, Goedbloed09, Goedbloed10}. The equation of motion for the Lagrangian displacement $\tmmathbf{\xi}_1$ is  
\begin{equation}\label{eq:force_op}
\rho_0  \frac{\partial^2 \tmmathbf{\xi}_1}{\partial t^2} + 2 \rho_0 \left( \tmmathbf{v}_0 \cdot \nabla
\right)\frac{\partial \tmmathbf{\xi}_1}{\partial t} - \tmmathbf{G} \left[ \tmmathbf{\xi}_1\right]=0,
\end{equation}
 where the generalized force operator $\tmmathbf{G}$ is given by
\begin{multline*}
\tmmathbf{G} \left[ \tmmathbf{\xi}_1 \right] =
\rho_0\left(\frac{m}{r}v_{0\varphi}+kv_{0z}
\right)^2\tmmathbf{\xi}_1-2i\rho_0\Omega\left(\frac{m}{r}v_{0\varphi}+kv_{0z}\right)\left(\xi_{1\varphi}{\hat{\bf
r}}-\xi_{1r}\hat{\tmmathbf{
\varphi}}\right)-\\
-2\rho_0r\Omega\frac{d\Omega}{dr}\xi_{1r}\hat{\bf
r}+\rho_1(r\Omega^2)\hat{\bf r}+{\tmmathbf
J}_0\times {\tmmathbf B}_1+(\nabla\times {\tmmathbf B}_1)\times
{\tmmathbf B}_0,
\end{multline*}
and the $0$ subscript indicates again the equilibrium quantities, while the $1$ subscript indicates perturbations. Substituting  $\tmmathbf{\xi}_1 \propto \exp({\rm i} \omega t )$ in equation (\ref{eq:force_op}), we get
\begin{equation}\label{eq:var_disprel}
\rho_0\omega^2 \tmmathbf{\xi}_1-2{\rm i} \rho_0\omega \left( \tmmathbf{v}_0 \cdot \nabla
\right) \tmmathbf{\xi}_1+ \tmmathbf{G} \left[ \tmmathbf{\xi}_1\right]=0.
\end{equation}
The various terms entering the expression of $\tmmathbf{G}$ correspond to different forces acting on the perturbations in the jet flow. The first term comes from a convective derivative and describes the advection of perturbations by the mean flow. In unmagnetized flows, this term contributes to the KH instability. The second term is related to Coriolis force due to rotation, the third term is related to shear, or differential rotation of the flow, since it is proportional to the radial derivative of the angular velocity $\Omega$, the fourth term proportional to $\rho_1$ corresponds to the centrifugal force (radial buoyancy), the fifth and sixth terms are the linearized Lorentz force, respectively, due to the equilibrium current ${\tmmathbf J}_0$ and the perturbed magnetic field ${\tmmathbf B}_1$ and due to the perturbed current $\nabla\times {\tmmathbf B}_1$ and the equilibrium magnetic field
${\tmmathbf B}_0$.

One can show that the force operator $\tmmathbf{G}$ is self-adjoint
\[ \int \tmmathbf{\eta} \cdot \tmmathbf{G} \left[ \tmmathbf{\xi}_1 \right] d^3
   \tmmathbf{r}= \int \tmmathbf{\xi} _1\cdot \tmmathbf{G} \left[ \tmmathbf{\eta}
   \right] d^3 \tmmathbf{r} \]
while the second term in equation (\ref{eq:var_disprel}) is antisymmetric
\[ \int \rho_0 \tmmathbf{\eta} \cdot \left( \tmmathbf{v}_0 \cdot \nabla \right)
   \tmmathbf{\xi}_1d^3 \tmmathbf{r}= - \int \rho_0 \tmmathbf{\xi}_1 \cdot \left(
   \tmmathbf{v}_0 \cdot \nabla \right) \tmmathbf{\eta}d^3 \tmmathbf{r}, \]
where $\tmmathbf{\eta}$ is an arbitrary function and integration is performed over an entire fluid volume provided that displacement $\tmmathbf{\xi}_1$ and $\tmmathbf{\eta}$ vanish at the flow boundaries. If we take $\tmmathbf{\eta}=\tmmathbf{\xi}^{\ast}_1$, we can write
\[ \int \tmmathbf{\xi}^{\ast}_1 \cdot \tmmathbf{G} \left[ \tmmathbf{\xi}_1 \right]
   d^3 \tmmathbf{r}= \int \tmmathbf{\xi}_1 \cdot \tmmathbf{G} \left[
   \tmmathbf{\xi}^{\ast}_1\right] d^3 \tmmathbf{r} \]
and
\[ \int \rho_0 \tmmathbf{\xi}^{\ast}_1 \cdot \left( \tmmathbf{v}_0 \cdot \nabla
   \right) \tmmathbf{\xi}_1d^3 \tmmathbf{r}= - \int \rho_0 \tmmathbf{\xi}_1 \cdot
   \left( \tmmathbf{v}_0 \cdot \nabla \right) \tmmathbf{\xi}^{\ast}_1 d^3
   \tmmathbf{r} \]
therefore $\int \tmmathbf{\xi}^{\ast}_1 \cdot \tmmathbf{G} \left[ \tmmathbf{\xi}_1
\right] d^3 \tmmathbf{r}$ is a real quantity and $\int \rho_0
\tmmathbf{\xi}^{\ast}_1 \cdot \left( \tmmathbf{v}_0 \cdot \nabla \right)
\tmmathbf{\xi}_1d^3 \tmmathbf{r}$ purely imaginary. We will see below that these properties are necessary for establishing stability criteria for the flow. 

Multiplying equation (\ref{eq:var_disprel}) by $\tmmathbf{\xi}^{\ast}_1$, integrating by $r$ over the interval
$[0, \infty]$ and taking into account that the perturbations vanish for $r\rightarrow \infty$ and are regular at $r=0$, we get
\begin{equation}
 A \omega^2 - 2 E \omega + F = 0, \label{eig3}
\end{equation}
where the coefficients $A, E$ and $F$ are  
\begin{eqnarray}\label{eq:var1}
  A & = & \int^{\infty}_0 \rho_0 \left| \tmmathbf{\xi}_1 \right|^2 r d r \nonumber \\
  E & = & \int^{\infty}_0 \left[ \rho_0 \left(\frac{m}{r} v_{0\varphi}+kv_{0z}\right) \left| \tmmathbf{\xi}_1
  \right|^2 + i \rho_0\Omega\left(\xi_{1r} {\xi}_{1\varphi}^{\ast} - \xi_{1\varphi} \xi^{\ast}_{1r} \right) \right] r d r \nonumber \\
  F & = & \int^{\infty}_0 \tmmathbf{\xi}^{\ast}_1 \cdot \tmmathbf{G} \left[
  \tmmathbf{\xi}_1 \right] r d r.  
\end{eqnarray}
$A$ and $E$ are real by definition, while $F$ is real due to the
self-adjointness of the force operator ${\tmmathbf G}$.
Using the expression of $\tmmathbf{G}$ in equation (\ref{eq:var1}), we can 
write $F$ in a symmetric form with respect to $\tmmathbf{\xi}_1$ and
$\tmmathbf{\xi}^{\ast}_1$:
\begin{multline}\label{eq:force_op_sym}
F= \int^{\infty}_0 \left[ \rho_0 \left(\frac{m}{r}
v_{0\varphi}+kv_{0z}\right)^2 \left| \tmmathbf{\xi}_1
  \right|^2 -2 i \rho_0\Omega \left(\frac{m}{r}v_{0\varphi}+kv_{0z}\right)\left(\xi_{1\varphi}\xi_{1r}^{\ast} -
  \xi_{1r}\xi_{1\varphi}^{\ast}\right)\right.-\\
  \left.
  -2\rho_0r\Omega\frac{d\Omega}{dr}|\xi_{1r}|^2+\frac{1}{2}(r\Omega^2)(\rho_1\xi_{1r}^{\ast}+\rho_1^{\ast}\xi_{1r})
  - \frac{1}{2} \tmmathbf{J}_0 \cdot \left( \tmmathbf{\xi}_1^{\ast} \times
   \tmmathbf{B}_1+\tmmathbf{\xi}_1 \times \tmmathbf{B}_1^{\ast} \right) - |\tmmathbf{B}_1|^2\right] r dr, 
\end{multline}
where  the various terms are grouped according to driving forces they correspond to, as in $ {\tmmathbf G}$.
The solution to the quadratic equation (\ref{eig3}) is
\begin{equation}\label{eq:omg_var}
\omega = \frac{E \pm \sqrt{E^2 - AF}}{A}. 
\end{equation}
Of course this is a formal solution, since the terms $A$, $E$ and $F$ depend on the eigenfunctions, so they can be computed only after the eigenvalue problem has been solved. If $E^2 < AF$ for an eigenmode, this solution comes in complex conjugate pairs that indicates instability of the mode. Therefore, in the expression (\ref{eq:force_op_sym}) for $F$, negative terms are stabilizing and positive ones destabilizing.  We distinguish four distinct destabilizing contributions:
\begin{enumerate}
\item
The sum of the first two terms 
\begin{equation}\label{eq:F1}
F_1= \int^{\infty}_0 \left[ \rho_0 \left(\frac{m}{r}
v_{0\varphi}+kv_{0z}\right)^2 \left| \tmmathbf{\xi}_1
  \right|^2 -2 i \rho_0\Omega \left(\frac{m}{r}v_{0\varphi}+kv_{0z}\right)\left(\xi_{1\varphi}\xi_{1r}^{\ast} -
  \xi_{1r}\xi_{1\varphi}^{\ast}\right) \right]r dr 
\end{equation}
describe the combined effect of advection by the mean flow and Coriolis force. However, these processes also define $E$ and only the sign of $F'_1\equiv E^2-AF_1$ actually characterizes stabilizing or destabilizing contribution due to these two effects.  This term is responsible for the velocity shear, or KH instability.
\item
The third term
\begin{equation}\label{eq:Fsh}
F_{sh} = -2\int^{\infty}_0 \rho_0r\Omega\frac{d\Omega}{dr}|\xi_{1r}|^2 r dr 
\end{equation}
describes the effect of shear, or differential rotation and is destabilazing when $F_{sh}>0$, i.e., $d\Omega/dr<0$ somewhere in the flow field. $F_{sh}$, together with the combined effect of advection and Coriolis force characterized by $F'_1$, determines instability in shear flows. In magnetized shear flows, the condition $d\Omega/dr<0$ is necessary for the existence of the MRI \citep{Balbus92}. 
\item
The fourth term 
\begin{equation}\label{eq:Fc}
F_c=   \frac{1}{2}\int^{\infty}_0(r\Omega^2)(\rho_1\xi_{1r}^{\ast}+\rho_1^{\ast}\xi_{1r}) r dr,
\end{equation}
which is proportional to the centrifugal acceleration, describes the effect of centrifugal force.  If $F_c>0$, the centrifugal force can give rise to the magnetic buoyancy instability.  This term depends on the density perturbation, which is expressed via $\xi_{1r}$ and $P_1$ as
\begin{equation}
\rho_1=\frac{\rho_0}{B_0^2}\left[P_1-(2B_{0\varphi}^2\tilde{\omega}^2+2v_{0\varphi}B_{0\varphi}k_B\tilde{\omega}-v_{0\varphi}^2(\rho_0\tilde{\omega}^2-k_B^2))\frac{\xi_{1r}}{r\tilde{\omega}^2}\right]
\end{equation}
At large $k$, to leading order the density perturbation becomes 
\[
\rho_1\approx -\frac{\rho_0}{rB_0^2\tilde{\omega}^2}v_{0\varphi}^2k^2B_{0z}^2\xi_{1r},
\] 
indicating that it is produced mainly by bending poloidal field lines. By contrast, at small $k$, $k_B \approx mB_{0\varphi}/r$ and the density perturbation is determined primarily by bending the toroidal field (especially at small pitch, when the growth rates are higher). This implies that the density perturbation in this regime arises due to bending of toroidal field lines.
 \item
The fifth term
\begin{equation}\label{eq:Fcd}
F_{cd}=  - \frac{1}{2}\int^{\infty}_0 \tmmathbf{J}_0 \cdot \left(\tmmathbf{\xi}_1^{\ast} \times
   \tmmathbf{B}_1+\tmmathbf{\xi}_1 \times \tmmathbf{B}_1^{\ast} \right) r dr
\end{equation}
is proportional to the equilibrium current and corresponds to the Lorentz force. If $F_{cd}>0$, this term is destabilizing, giving rise to current driven instability. The last term is the magnetic tension force, which is always stabilizing. 
\end{enumerate}
Based on the above analysis, we classify unstable modes according to which of these four contributions prevails over the net effect of other three ones and results in the destabilization of a given mode. We then label the mode according to the type of this dominant destabilizing term. So, for example, if  $F_{sh}$ is positive and dominates over the net contribution from all the other terms in the square root in equation (\ref{eq:omg_var}), this implies that the instability is caused by differential rotation, which in the case of the considered jet flow threaded by the magnetic field in fact corresponds to the ``cold'' version of MRI. If $F_c$ is positive and dominates, the main destabilizing force is the centrifugal force, which via bending magnetic field lines, gives rise to the magnetic buoyancy instability. Finally, if $F_{cd}$ term is positive and dominates, the destabilization comes from the Lorentz force due to the presence of the equilibrium current and hence the resulting instability is current driven.


\section{Results}
\label{results}
%
%

\begin{figure}
   \centering
   \includegraphics[width=15cm]{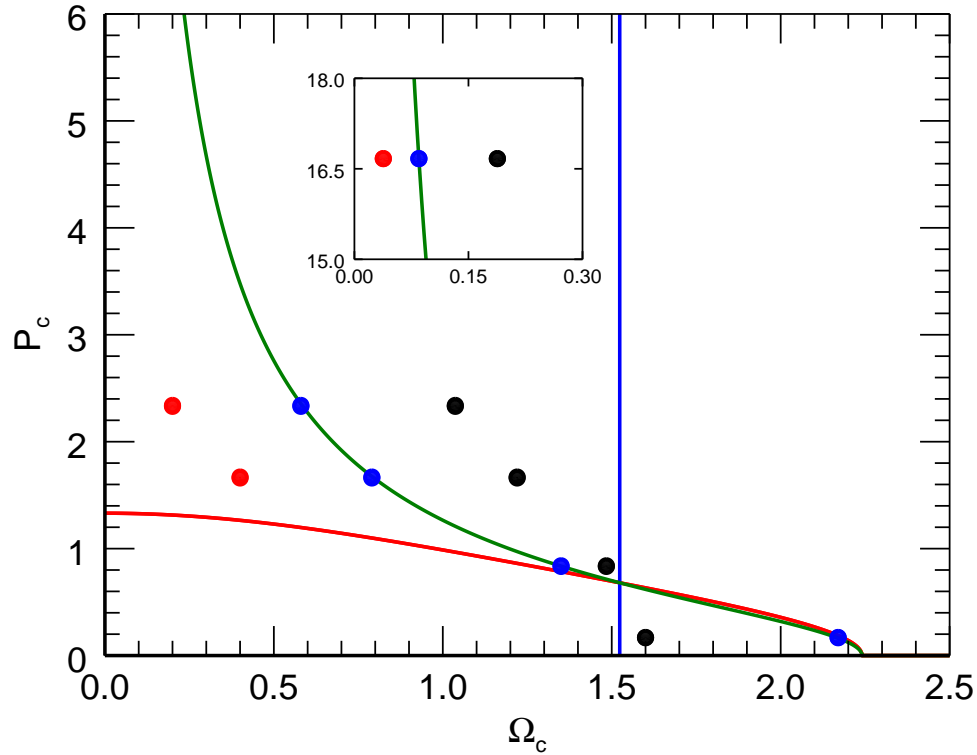} 
   \caption{\small  The equilibrium solutions in the parameter plane $(\Omega_c,P_c)$ for which we computed the behavior of unstable modes. The red dots represents equilibria with $\alpha = 0.2$, the blue dots are for $\alpha = 1$ and the black dots are for $\alpha = 5$. The green curve corresponds to solutions with $\alpha = 1$.}
   \label{fig:figcases}
\end{figure}

As discussed in Section \ref{sec:equilibrium}, the basic equilibrium depends on the two parameters $\Omega_c$ and $P_c$, wich are  defined in equation (\ref{eq:param}) and represent respectively nondimensional measures of the rotation rate  and of the pitch on the vertical axis. Instead of $\Omega_c$,  which measures the rotation rate in terms of the average total Alfv{\' e}n velocity,  we can alternatively make use of the parameter $\alpha$, defined in equation (\ref{eq:alfa}), which measures rotation in terms of the Alfv\'en velocity associated only with the azimuthal magnetic field component. Reference to parameter $\alpha$ can be convenient  because its value can be related to the sign of the radial gradient of $B_z$, i.e. for $\alpha < 1$, $B_z$ decreases outward, for $\alpha = 1$, $B_z$ is constant and, for $\alpha > 1$, $B_z$ increases outward.  In the following we will focus our discussion on a number of equilibrium solutions, whose position in the $(\Omega_c, P_c)$ plane is shown in Fig. \ref{fig:figcases}. The green curve corresponds to solutions with $\alpha = 1$, the red dots represents equilibria with $\alpha = 0.2$, the blue dots are for $\alpha = 1$ and, finally, the black dots are for $\alpha = 5$. We choose these three values of $\alpha$ in order to sample the solutions with different gradients of $B_z$.
\subsection{Axisymmetric modes $(m=0)$}
\label{subsec:axisym}
\begin{figure}
   \centering
   \includegraphics[width=15cm]{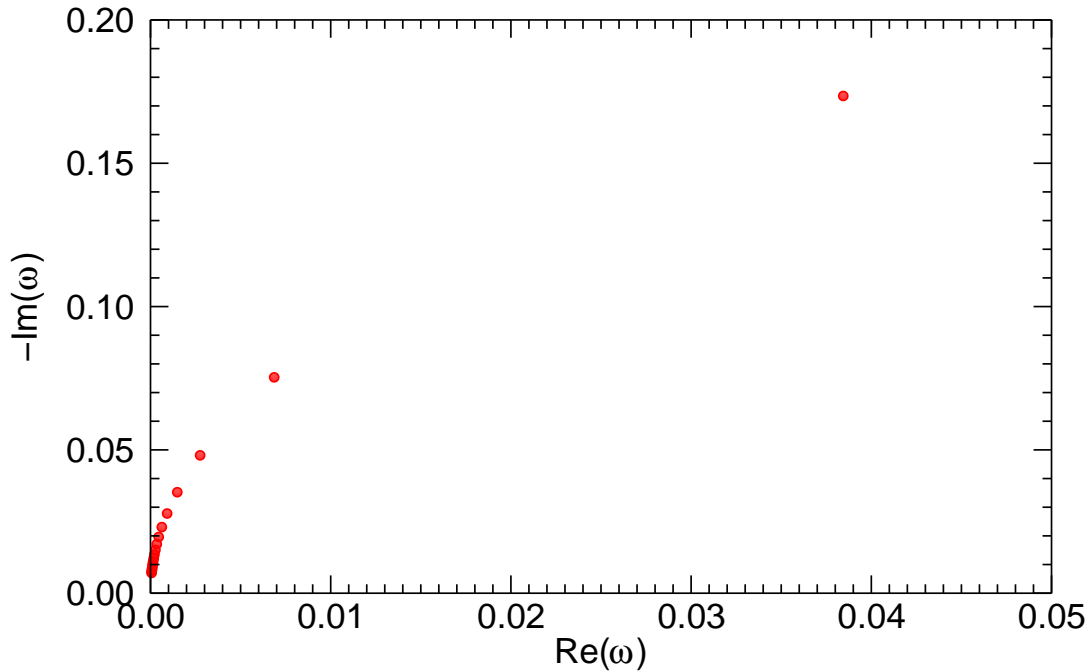} 
   \caption{\small  Location of unstable modes in the complex plane. The values of the parameters are $\alpha = 1$, $P_c = 1.66$ (corresponding to $\Omega_c=0.79$), $m=0$ and $ka=1.5$.  }
   \label{fig:complexplanem0}
\end{figure}

We start our discussion with the axisymmetric modes, in this case we know that the CDI mode is stable and instabilities can be only due to rotation. In Fig. \ref{fig:complexplanem0} we plot in the complex plane the position of unstable modes for a given parameter set $P_c = 1.66$,   $\Omega_c = 0.79$ ($\alpha=1$) and $ka=1.5$. We observe a sequence of modes clustering to $\omega=0$, that is the point where the slow continuum collapses in the present conditions ($p_0 = 0$, $v_{0z} = 0$ and $m=0$). The modes  in the sequence differ by the number of radial oscillations which increases as the sequence approaches $\omega =0$.

\begin{figure}
   \centering
   \includegraphics[width=15cm]{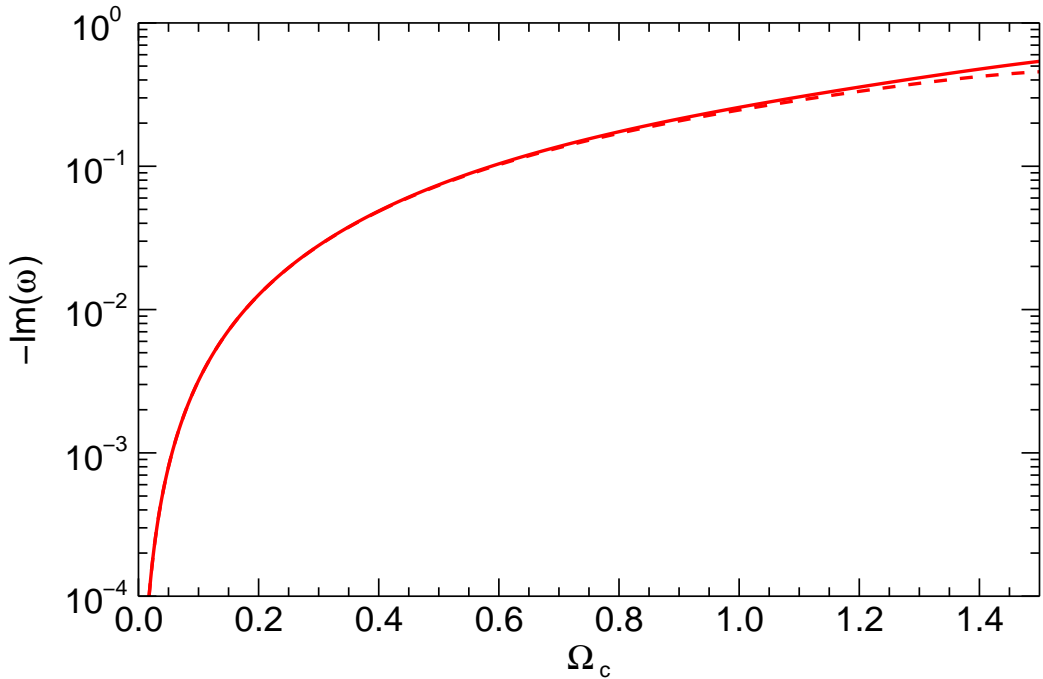} 
   \caption{\small  Plot of the growth rate versus the rotation rate $\Omega_c$ for the most unstable mode of the axisymmetric centrifugal buoyancy mode shown in Fig. \ref{fig:complexplanem0}. The dashed curve represents an approximation to the growth rate given by equation (\ref{eq:cengr}). }
   \label{fig:centrifugalm0}
\end{figure}

We can then investigate the physical origin of this sequence of modes  by comparing, in Fig. \ref{fig:centrifugalm0}, the growth rate of the most unstable one as a function of $\Omega_c$ (solid curve; the value of $ka$ is again 1.5) with an approximation obtained by equation (\ref{eq:omg_var}) in which we consider only  the centrifugal term (dashed curve), i.e. we approximate the growth rate by
\begin{equation}\label{eq:cengr}
-\Im(\omega) = \sqrt{\frac{F_c}{A}},
\end{equation}
where $F_c$ and $A$ are given respectively by equations (\ref{eq:Fc}) and (\ref{eq:var1}) and can be computed once we have solved the eigenvalue problem and found the eigenfunctions. We can see that the approximation reproduces very well the behavior of the actual growth rate, so we can regard the centrifugal term as being responsible for the destabilization of these modes and hence identify them as magnetic buoyancy instabilities. We already discussed these instabilities in Section \ref{modeclassification}, when we considered the WKBJ local dispersion relation. They have been also already studied by \citet{Huang03} and \citet{Kim00} and, as we mentioned above, are analogous to the Parker instability, with the centrifugal force replacing gravity and operate by bending the poloidal field lines. 

%

\begin{figure}
   \centering
   \includegraphics[width=15cm]{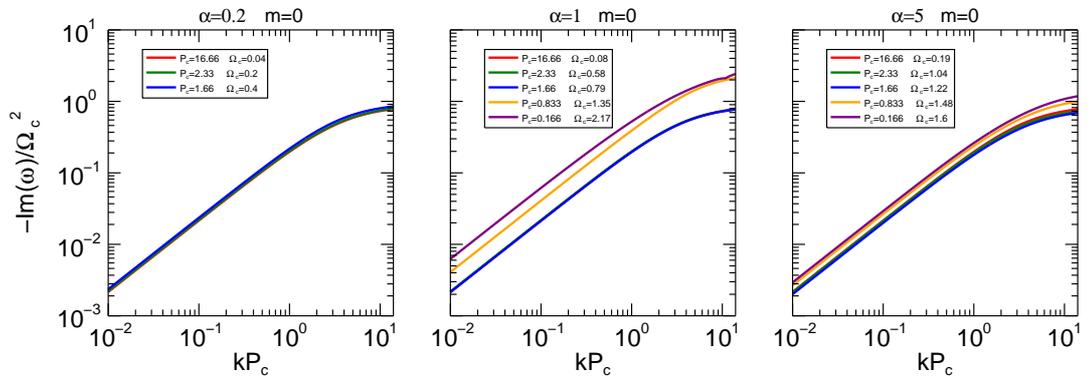} 
   \caption{\small  Plot of the growth rate versus the wavenumber for axisymmetric modes. The three panels refer to three different values of $\alpha$, the left panel is for $\alpha = 0.2$, the mid panel is for $\alpha=1$ and the right panel is for $\alpha=5$. The different curves refer to different values of $P_c$ and the corresponding values of $P_c$ and $\Omega_c$ are given in the legend.}
   \label{fig:m0}
\end{figure}

From the local dispersion relation (\ref{eq:pbm}) appropriate for the poloidal buoyancy mode, we expect their growth rates to be proportional to the square of the rotation rate. In Fig. \ref{fig:m0} we then show their growth rates divided by $\Omega_c^2$ as a function of the wavenumber. The three panels are for three different values of $\alpha$ and the different curves in each panel are for different values of $P_c$. As already discussed, the centrifugal buoyancy modes represent actually a sequence of unstable modes and, in the panels, we show only those branches of this mode with the maximum growth rate. This figure demonstrates that  the $\Omega_c^2$ scaling law is quite good and that, as expected from the local dispersion relation,  the growth rate increases with the vertical wavenumber and tends to an asymptotic limit as $k \rightarrow \infty$, in fact the behaviour of the growth rate as a function of the wavenumber is the same as in Fig. \ref{fig:figwkbj}. These modes appear to be always unstable, the reason, discussed by \citet{Huang03},  is related to the fact that, since the plasma has no pressure, it is possible to compress it  along the field lines and create a density perturbation without performing any work. The centrifugal force can then always overcome the magnetic restoring forces. However, the inclusion of a finite pressure tends to stabilize this mode.   

\subsection{Non-axisymmetric modes $(m \neq 0)$}
\label{subsec:nonaxisym}
\begin{figure}
   \centering
   \includegraphics[width=15cm]{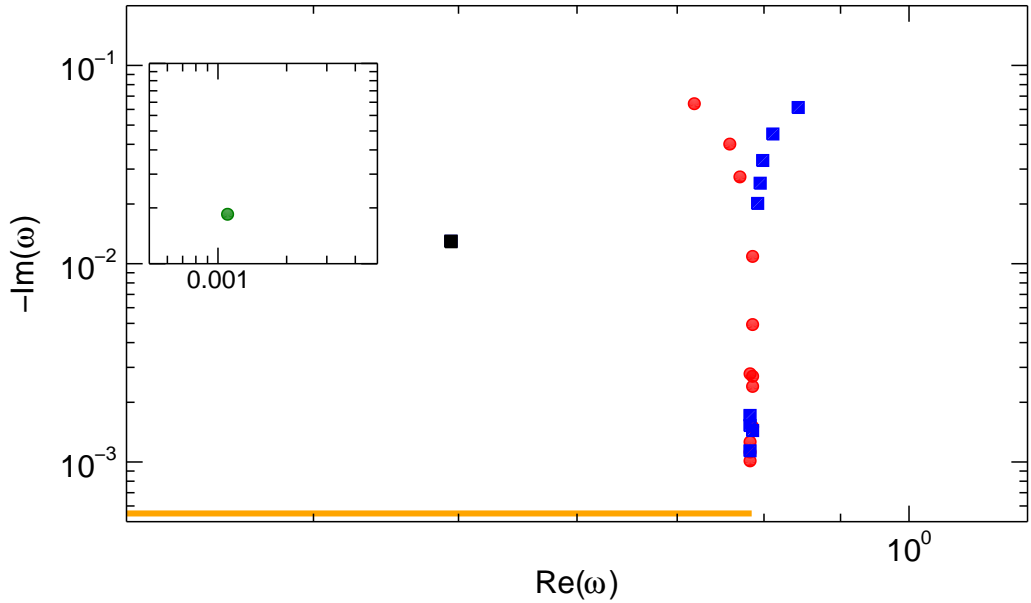} 
   \caption{\small  Location of unstable modes in the complex plane. The values of the parameter are $\alpha = 1$, $P_c = 1.66$, $m=1$, the (green and red) dots are for $ka=0.12$, while the (black and blue) squares are for $ka=0.72$. The orange line shows the frequency range of the slow continuum, which, in the zero pressure case, reduces to the flow continuum. As it is discussed in the text, the green dot represents the CDI, the black dot MRI, while the red and blue series represent centrifugal modes.  } 
   \label{fig:complexplanem1}
\end{figure}

\begin{figure}
   \centering
   \includegraphics[width=15cm]{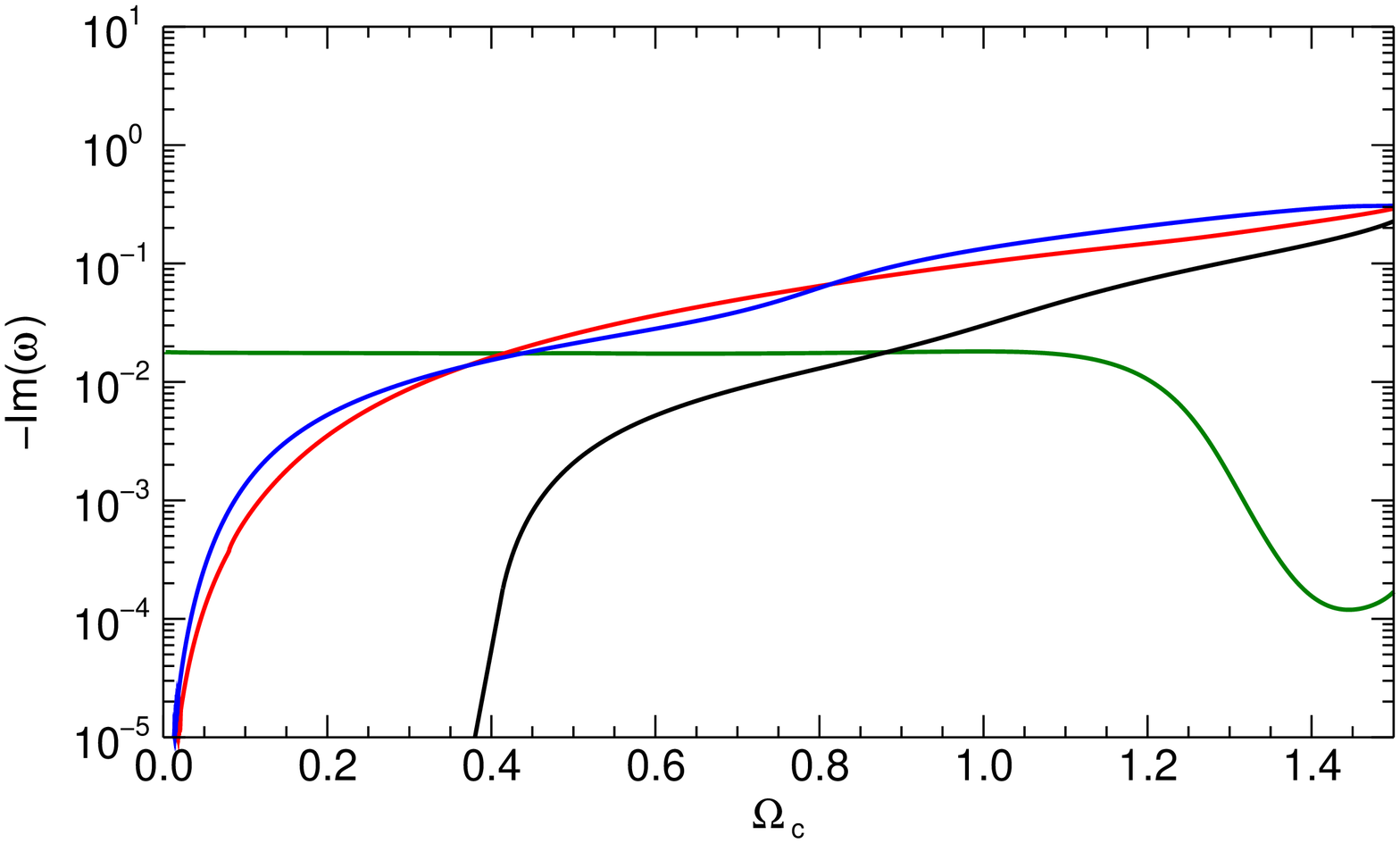} 
   \caption{\small Plot of the growth rate versus the rotation rate $\Omega_c$ for the modes represented in Fig. \ref{fig:complexplanem1}. The colors of the curves correspond to the color in  Fig. \ref{fig:complexplanem1}. }
   \label{fig:grvsalfa}
\end{figure}

%
%
We start our analysis of non-axisymmetric instabilities by considering first the cases with $\alpha = 1$ and $m = 1$ and, in order to get a first indication on the number and the kind of modes that we can find, in Fig. \ref{fig:complexplanem1} we plot in the complex plane the position of unstable modes for a given parameter set.  The figure is for $P_c = 1.66$ and we consider two values of  the wavenumber: squares are for $ka=0.72$ and dots  are for $ka = 0.12$. Both at large and small wavenumbers, we observe an isolated mode  and a series. In Fig. \ref{fig:grvsalfa} we consider the behavior of the modes represented in Fig. \ref{fig:complexplanem1} as a function of the rotation parameter $\Omega_c$, the colors of the curves are in correspondence with the colors in  Fig. \ref{fig:complexplanem1}, and, for the series, we have considered only the mode with the largest growth rate.  From the figure we can see that the only mode that survives when we let rotation go to zero  is the one  corresponding to  the green dot in Fig. \ref{fig:complexplanem1}, all the others become stable. We can then conclude that the mode corresponding to the green dot reduces to the CDI mode in the zero rotation limit, while rotation is at the origin of all the other modes. In this figure, we can further notice a stabilizing effect of rotation on the CDI, however, this is noticeable only at large values of $\Omega_c$. This can be compared with the results of \citet{Carey09} who find also a stabilizing effect of rotation, but at smaller values of the rotation rate, in their case however there is a rigid rotation, while in this case the rotation rate decreases radially.

\begin{figure}
   \centering
   \includegraphics[width=15cm]{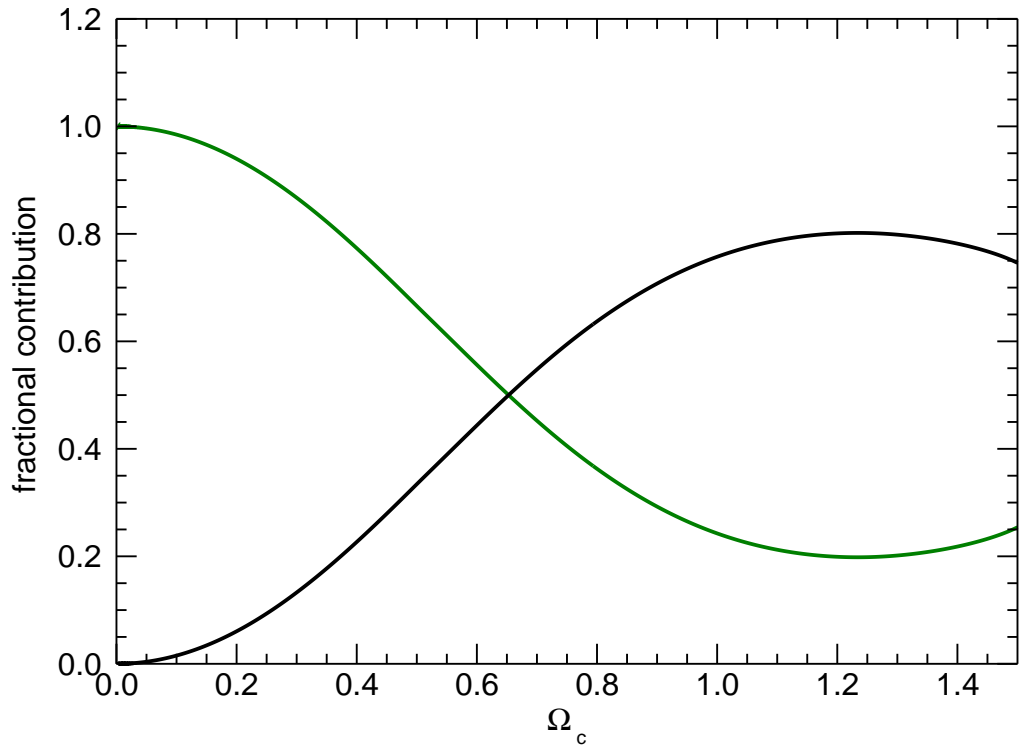} 
  \caption{\small Plot of the fractional contributions of the current term, $f_{cd}$ (green), and of the shear term, $f_{sh}$ (black), as a function of the rotation rate $\Omega_c$ for the CDI mode. The values of other parameters are the same as in Fig. \ref{fig:complexplanem1}. } 
   \label{fig:cdvar}
\end{figure}

We can further investigate the physical origin of the different modes by computing the stabilizing and destabilizing terms based on the energetic considerations discussed in the previous section. Starting from the CDI mode, we find that the destabilizing terms for this mode is not only $F_{cd}$, as it should be expected for the current driven mode, but also $F_{sh}$. In Fig. \ref{fig:cdvar} we plot the fractional contributions of these two terms as a function of the rotation rate $\Omega_c$. The fractional contributions for the two terms are defined, respectively, as
\begin{equation}\label{eq:fraccd}
f_{cd} =  F_{cd}/(F_{cd}+F_{sh})
\end{equation}
and 
\begin{equation}\label{eq:fracsh}
f_{sh} = F_{sh}/(F_{cd}+F_{sh})
\end{equation}
We see that at low rotation rates, the dominant term is the current term represented by the green curve, but, as we increase the rotation rate, the contribution by  the shear term increases until it becomes dominant for  $\Omega_c > 1$. 

\begin{figure}
   \centering
   \includegraphics[width=15cm]{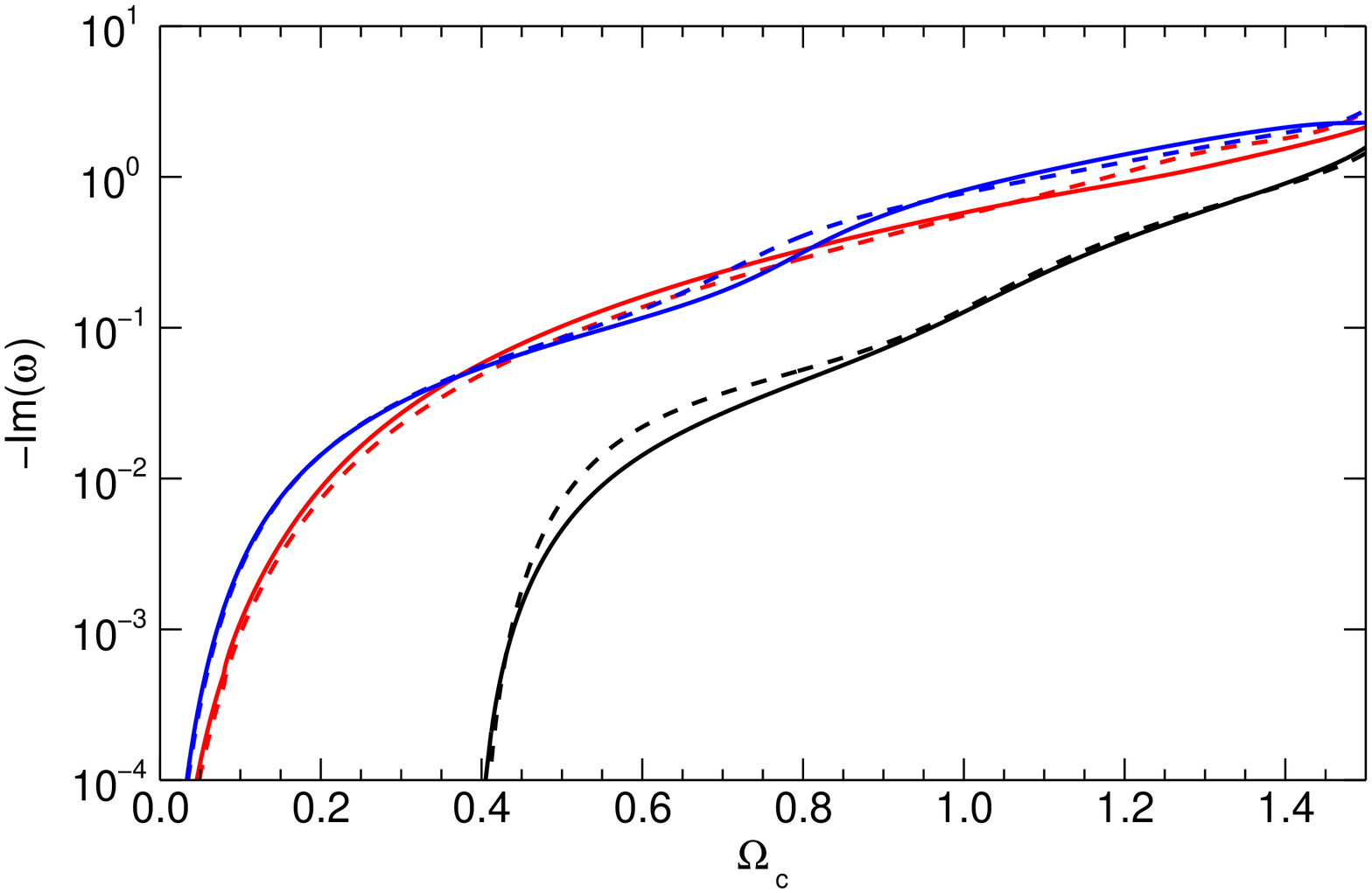} 
   \caption{\small  Plot of the growth rate vs. the rotation rate $\Omega_c$ for the centrifugal buoyancy modes (blue and red) and the MRI branch (black). The blue and red dashed curves represent an aproximation to the growth rate given by equation (\ref{eq:cengr}) for the buoyancy modes, whereas the black dashed curve represents an approximation to the growth rate given by equation (\ref{eq:growthMRI}) for the MRI. The values of other parameters are the same as in Fig. \ref{fig:complexplanem1}.}
 \label{fig:centrifugal}
\end{figure}

As we did for the axisymmetric modes, we can investigate the physical origin of the two series of modes, represented by blue squares and red dots in  Fig. \ref{fig:complexplanem1},  by comparing in Fig. \ref{fig:centrifugal} the growth rate as a function  of $\Omega_c$ (solid curves) with the approximation given by equation (\ref{eq:cengr}) (dashed curves). We  see that the approximation reproduces fairly well the behavior of the actual growth rate, so we can then regard the  centrifugal term as being responsible for the destabilization of these modes and hence identify them as magnetic buoyancy instabilities. As discussed in the previous section and in Section \ref{sec:wkb}, with the WKBJ analysis, we can distinguish them as a toroidal buoyancy mode at low wavenumbers (red curve) and a poloidal buoyancy mode at high wavenumbers (blue curve). The two sequences of modes cluster to the edge of the flow continuum (which is what is left of the slow continua in the zero pressure case), whose frequency range is represented in Fig. \ref{fig:complexplanem1} by the orange line.

Consider now the mode represented by  the black square in Fig. \ref{fig:complexplanem1}. The black curve in Fig. \ref{fig:centrifugal} traces this mode as $\Omega_c$ varies. Similarly to the buoyancy modes, we can aproximate its growth rate by an expression analogous to equation (\ref{eq:cengr}) in which $F_c$ is replaced by $F_{sh}$, 
\begin{equation}\label{eq:growthMRI}
-\Im(\omega) = \sqrt{\frac{F_{sh}}{A}}.
 \end{equation}
The dashed black curve in Fig. \ref{fig:centrifugal} shows the growth rate given by equation (\ref{eq:growthMRI}), which indeed closely follows an actual one represented by the black curve. So, the main driving force for this mode is related to the shear of the radially decreasing rotation rate and therefore it should be identified with the MRI.  However, this occurs only in a very limited parameter range, while in other regions it merges with either the buoyancy modes or with the CDI mode, where the driving force become either the centrifugal term or the current term, respectively (see below). So, from now on we mostly concentrate on the CDI and centrifugal buoyancy modes.

\begin{figure}
   \centering
   \includegraphics[width=15cm]{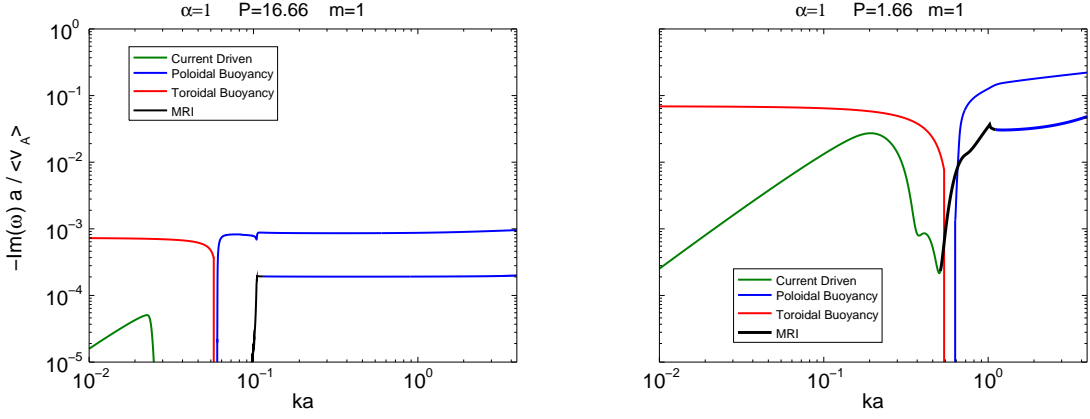} 
   \caption{\small Plot of the growth rate vs. the wavenumber for two cases with different values of the pitch parameter $P_c$. The left panel is for $P_c=16.66$ while the right panel is for $P_c = 1.66$. The other parameters are $\alpha = 1$ and $m=1$. }
   \label{fig:gr1}
\end{figure}

We can now proceed with a more detailed analysis of  the dependence of the growth rates  on the wavenumber, the pitch and rotation. In Fig. \ref{fig:gr1}, we plot the growth rates as a function of the wavenumber for $P_c = 16.66$ (left panel) and for $P_c=1.66$ (right panel) for $\alpha=1$ and $m=1$. In the left panel, we have clearly distinct the CDI and  the toroidal and poloidal buoyancy modes. For both values of the pitch, the black part of the curve, which corresponds to the MRI, is distinct only in the growing part over a relatively narrow range of wavenumbers and is merged with one of the poloidal  buoyancy modes in the constant region. As expected and discussed in paper I, the CDI mode (green curve) increases its growth rate and the value of its maximum unstable wavenumber as we decrease the pitch.  For $P_c \sim 1.66$, the increase of the maximum unstable wavenumber brings the CDI mode to a complicated interaction with the other modes, merging first with the MRI branch and then with one of the poloidal buoyancy modes. For $P_c \le1.66$  the CDI mode becomes  unstable for all wavenumbers, while the driving force, increasing the wavenumber, changes nature, becoming first related to the shear of rotation and then centrifugal. Both the  toroidal and the poloidal  buoyancy modes increase their growth rate as we decrease the pitch, this is because in the equilibrium configuration the rotation rate increases as the pitch decreases. The toroidal buoyancy mode becomes stable at high wavenumbers, while the poloidal buoyancy mode becomes stable at small wavenumbers. Decreasing $P_c$, the stable region between them shows a small increase in width and moves towards high wavenumbers.

\begin{figure}
   \centering
   \includegraphics[width=15cm]{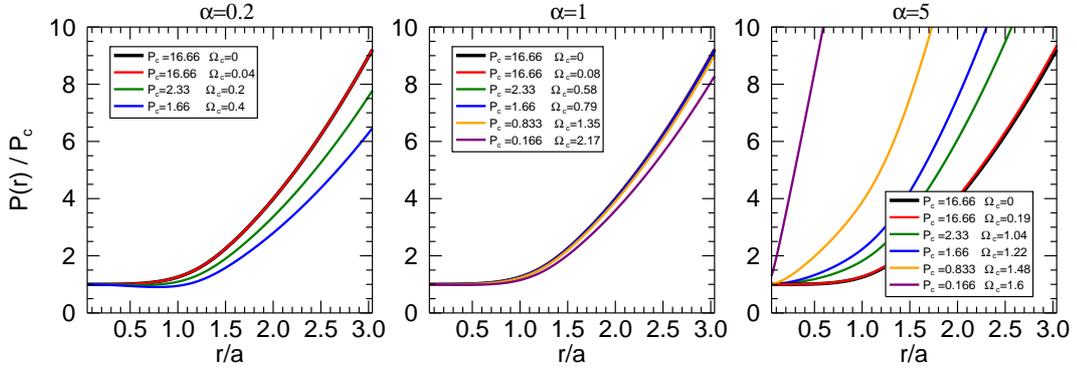} 
   \caption{\small Plot of the pitch as a function of radius for the equilibrium solutions shown in Fig. \ref{fig:figcases}. The three panels refer to three different values of $\alpha$, the left panel is for $\alpha = 0.2$, the mid panel is for $\alpha=1$ and the right panel is for $\alpha=5$. The different curves refer to different values of $P_c$ and the corresponding values of $P_c$ and $\Omega_c$ are given in the legend. }
   \label{fig:pitch}
\end{figure}

\begin{figure}
   \centering
   \includegraphics[width=15cm]{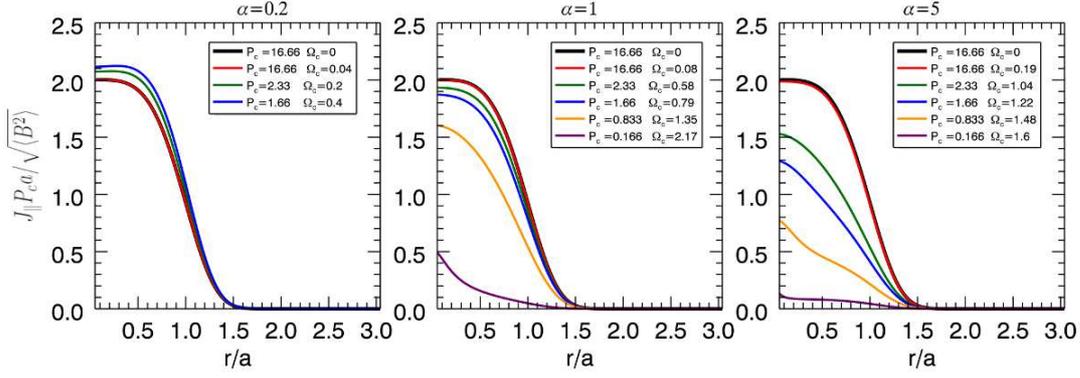} 
   \caption{\small Plot of the current component parallel to the magnetic field as a function of radius,   for the equilibrium solutions shown in Fig. \ref{fig:figcases}. The three panels refer to three different values of $\alpha$, the left panel is for $\alpha = 0.2$, the mid panel is for $\alpha=1$ and the right panel is for $\alpha=5$. The different curves refer to different values of $P_c$ and the corresponding values of $P_c$ and $\Omega_c$ are given in the legend. }
   \label{fig:jpar}
\end{figure}

For discussing in more detail the behavior of the CDI, it can be useful to examine how the  properties of the equilibrium structure are modified in the different parameter ranges. Figures \ref{fig:pitch} and \ref{fig:jpar} show the radial profiles of the pitch and the equilibrium current component parallel to the magnetic field, $J_{0,\parallel}$, which is the destabilizing factor for the CDI. In fact, by rearranging expression (\ref{eq:Fcd}) for the current term $F_{cd}$, one can show that the destabilizing term is due to a contribution proportional to the current component parallel to the magnetic field \cite[e.g., see][Ch. 8]{Freidberg87}. Therefore, the latter is a central quantity determining the growth rate of the CDI. The three panels are respectively for $\alpha = 0.2$ (left panel), $\alpha = 1$ (mid panel) and $\alpha = 5$ (right panel) and the different curves in each panel refer to different values of $P_c$, in each panel we also plotted the case with $P_c = 16.66$ and no rotation for reference (black curves). We remember that a variation of the pitch leads also to a variation of the rotation rate: for lower values of the pitch we have higher values of $\Omega_c$ and these values can be read in the legend. The pitch profile is normalized to the value $P_c$, on the axis,  while the  parallel current is multiplied by $P_c$ to bring curves for different values of $P_c$ on the same scale, since for large values of $P_c$ the parallel current scales as $1/P_c$. The pitch profile is in general characterized by a flat part up to $r=a$ followed  by a steep increase. In the left panel, we see that a decrease of $P_c$ leads to a slower increase of $P(r)$ for $r > a$. Correspondingly, in the left panel of Fig. \ref{fig:jpar}, we observe a slight increase of the parallel current.  In the case of $\alpha = 1$ (mid panel), when we have $B_z$ constant, the pitch profiles remain essentially unchanged when we decrease $P_c$, while  the parallel current shows a substantial decrease for low values of $P_c$. For $\alpha = 5$ (right panel),  $B_z$ increases with radius and consequently the pitch show increasingly steeper profiles as we decrease $P_c$ and the decrease of the  parallel current  is larger than in  the previous case.


\begin{figure}
   \centering
   \includegraphics[width=15cm]{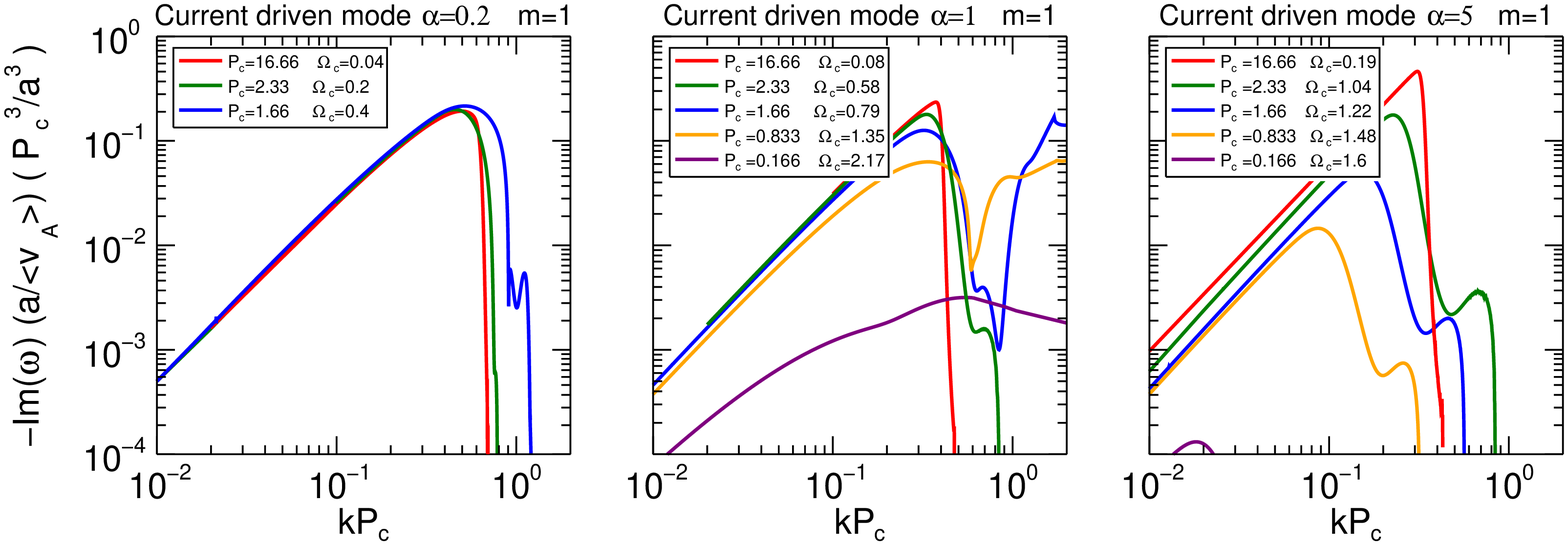} 
   \caption{\small  Plot of the growth rate of the CDI as a function of  $k P_c$ for $m=1$. The three panels refer to three different values of $\alpha$, the left panel is for $\alpha = 0.2$, the mid panel is for $\alpha=1$ and the right panel is for $\alpha=5$. The different curves refer to different values of $P_c$ and the corresponding values of $P_c$ and $\Omega_c$ are given in the legend. }
  \label{fig:cd}
\end{figure}

 In paper I we discussed a scaling law for  the growth rate of the  CDI of the form
\begin{equation}\label{eq:scaling}
-\Im(\omega) \sim \frac{ \langle v_A \rangle}{a} \left( \frac{a}{P_c} \right)^3 f(k P_c) 
\end{equation}
 and in Fig. \ref{fig:cd} we can investigate the effect of rotation by considering the deviation from  this scaling law. The three panels and the curves in each panel correspond to the same cases shown in Figs. \ref{fig:pitch} and \ref{fig:jpar}.  In the left panel we can observe  that a first effect of rotation is to move the cutoff wavenumber to smaller values, from $k P_c \sim 1$ without rotation (Paper I) to $k P_c \sim 0.8$. Apart from that, the scaling provided by Eq. (\ref{eq:scaling}) is quite good, slight deviations can be observed only for the smallest value of $P_c$ (blue curve), partly due to the interaction with other modes (MRI) and partly (as already discussed in Paper I) related to the  change of the pitch profile and parallel current observed in the corresponding equilibrium solution.  Comparing the red curves (largest values of $P_c$) in the three panels we see that the increase of the rotation rate leads the cutoff wavenumber to shift towards increasingly lower values, in parallel, however, we have also a slight increase of the growth rate in the unstable range. From the other curves (green, blue, orange and purple), we see that, for lower values of $P_c$,  at $\alpha = 1$,  the cutoff disappears because the CDI mode starts to interact and merge with the centrifugal mode and the growth rate decreases as a result of the decrease of the parallel current in the equilibrium configuration. For  $\alpha = 5$, we also observe  a decrease of the growth rate for increasingly lower values of  $P_c$, corresponding to the decrease of the parallel current. In summary, rotation has, in general,  a stabilizing effect on the CDI mainly because it modifies the equilibrium structure by decreasing the parallel current. This is consistent with \citet{Carey09}, who also found stabilization of CDI at high rotation rates in the case of rigid rotation. However, especially at large values of $P_c$, there are situations in which, on the contrary, the growth rate of CDI  shows a slight increase with the rotation rate. Finally, we recall that the CDI is stable for $m=-1$.

\begin{figure}
   \centering
   \includegraphics[width=15cm]{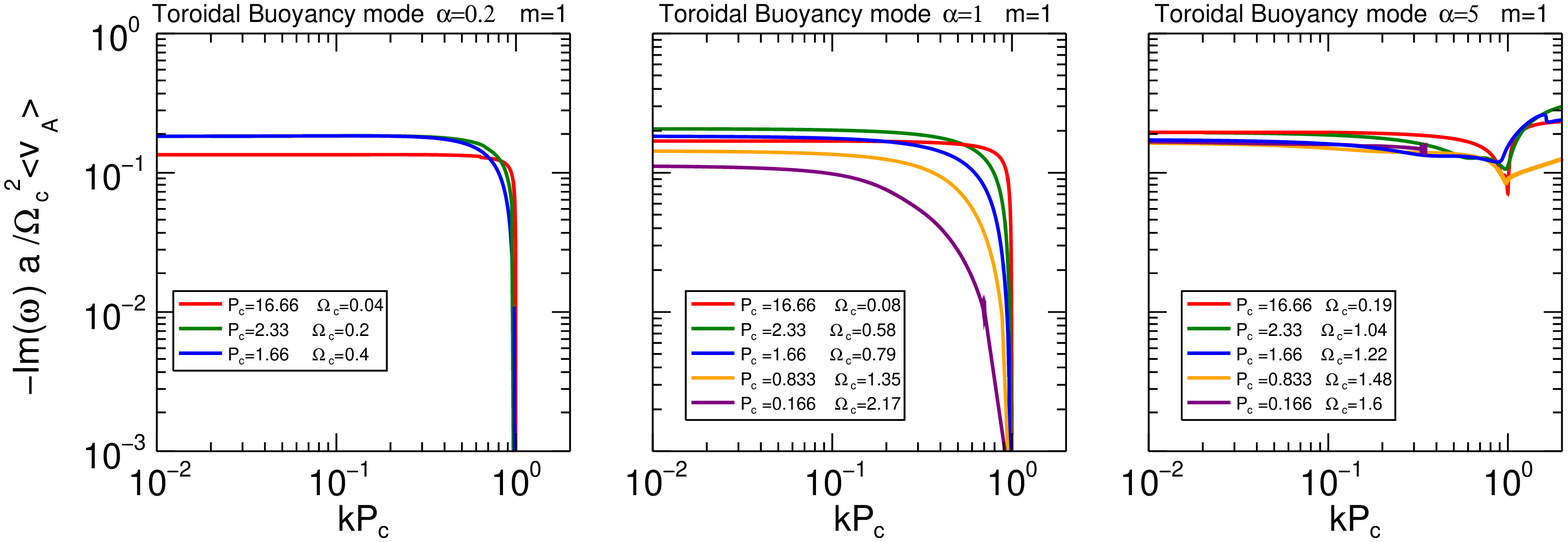} 
   \caption{\small Plot of the growth rate of the toroidal buoyancy mode as a function of  $k P_c$ for $m=1$. The three panels refer to three different values of $\alpha$, the left panel is for $\alpha = 0.2$, the mid panel is for $\alpha=1$ and the right panel is for $\alpha=5$. The different curves refer to different values of $P_c$ and the corresponding values of $P_c$ and $\Omega_c$ are given in the legend.  }
  \label{fig:tb}
\end{figure}

\begin{figure}
   \centering
   \includegraphics[width=15cm]{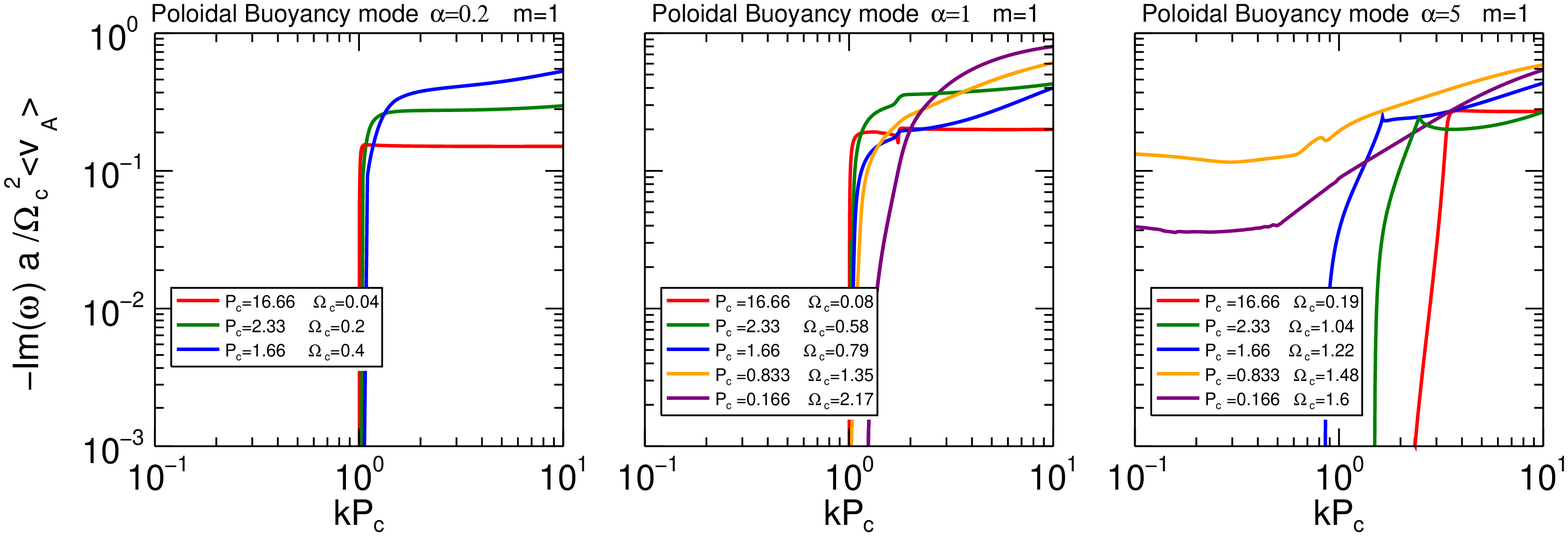} 
   \caption{ \small Plot of the growth rate of the poloidal buoyancy mode as a function of  $k P_c$ for $m=1$. The three panels refer to three different values of $\alpha$, the left panel is for $\alpha = 0.2$, the mid panel is for $\alpha=1$ and the right panel is for $\alpha=5$. The different curves refer to different values of $P_c$ and the corresponding values of $P_c$ and $\Omega_c$ are given in the legend. }
  \label{fig:pb}
\end{figure}

We can  now  turn our attention to  the toroidal and poloidal buoyancy modes. From the local dispersion relations, given by equations (\ref{eq:pbm}) and (\ref{eq:tbm}), we expect their growth rates to be proportional to the square of the rotation rate. In Figs. \ref{fig:tb} and \ref{fig:pb}, we then show their growth rates divided by $\Omega_c^2$ as a function of the wavenumber. The three panels are, as before, for three different values of $\alpha$ and the different curves in each panel are for different values of $P_c$. As already discussed, the toroidal and poloidal buoyancy modes represent actually a series of unstable modes and, in these figures, we show only the modes with the maximum growth rate. The $\Omega_c^2$ scaling law is quite good for the toroidal buoyancy mode, slightly less valid in the case of the poloidal buoyancy mode. In general, the value of $k P_c = 1$ represents a high wavenumber cutoff for the toroidal buoyancy mode and a low wavenumber cutoff for the poloidal buoyancy mode, however, for large values of the rotation rate, as shown in the right panels of Figs. \ref{fig:tb} and \ref{fig:pb}, we observe merging  between modes of the two series and deviations from the $k P_c = 1$ cutoff (note that the merging of poloidal and toroidal modes at $\alpha=5$ shown in the right panels of Figs. \ref{fig:tb} and \ref{fig:pb}, may refer to different mode branches in the two series).   

\begin{figure}
   \centering
   \includegraphics[width=15cm]{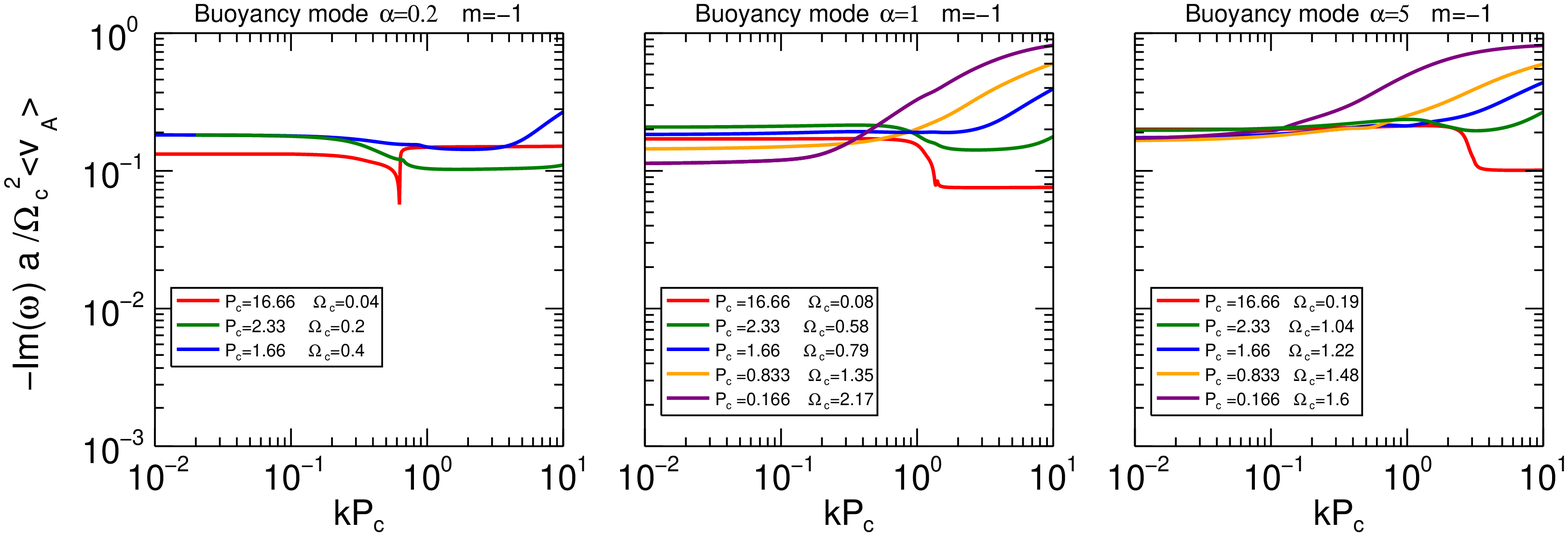} 

   \caption{ \small Plot of the growth rate of the centrifugal buoyancy mode as a function of  $k P_c$ for $m=-1$. The three panels refer to three different values of $\alpha$, the left panel is for $\alpha = 0.2$, the mid panel is for $\alpha=1$ and the right panel is for $\alpha=5$. The different curves refer to different values of $P_c$ and the corresponding values of $P_c$ and $\Omega_c$ are given in the legend. }
   \label{fig:buoym-1}
\end{figure}

 In Fig. \ref{fig:buoym-1} we show the results for the case with  $m=-1$. Overall, the buoyancy modes behave similarly with the wavenumber, although there are some differencies with the $m=1$ case at intermediate $k$. Specifically, we see that the buoyancy mode is unstable at all wavenumbers, in fact the narrow stability region around $kP_c=1$ disappears. The growth rate,  which is independent of the wavenumber at small values of the latter, shows however a variation around $k P_c = 1$, sometimes a decrease but typically an increase going towards larger values of the wavenumber. Of course, this increase of the growth rate eventually asymptotes to a finite value as $k\rightarrow \infty$, as in the $m=1$ case above, but at small $P_c$, this asymptotic value is reached at higher wavenumbers.   

\begin{figure}
   \includegraphics[width=\textwidth]{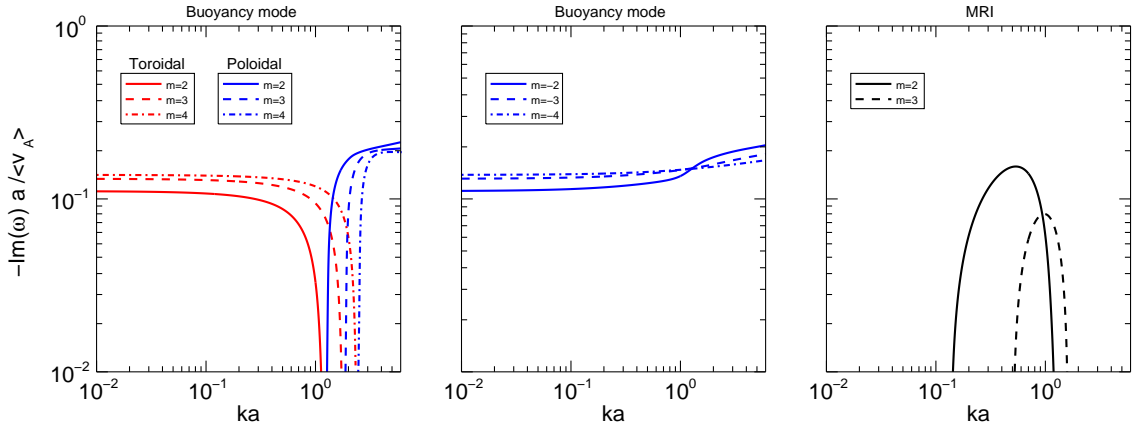} 
   \caption{\small Plot of the growth rate of the toroidal and poloidal buoyancy modes as well as the MRI at different $m=\pm 2$ (solid), $\pm 3$ (dashed), $\pm 4$ (dash-dot). The pitch parameter $P_c = 1.66$ and the rotation $\alpha = 1$.}
   \label{fig:diffm}
\end{figure}

So far we concentrated on the values $m=0,\pm 1$ of the azimuthal wavenumber and now we examine different values. Figure \ref{fig:diffm} shows the toroidal (red curves) and poloidal (blue curves) buoyancy modes as well as MRI (black curves) at different  $m$ and fixed $P_c=1.66$ and $\alpha=1$. First consider the case of positive $m$, when the toroidal and poloidal modes are separated. The growth rate of the toroidal mode increases with $m$ and the instability boundary extends to larger $k$. However, it is clear from this figure and also from the local dispersion relation (\ref{eq:tbmsol}) that the growth rate at small $k$ converges to a finite value as $m$ becomes large. By contrast, the growth rate of the poloidal mode decreases with $m$ and the instability boundary shifts to larger $k$, but the maximum growth achieved in the limit of high $k$ is essentially independent of $m$, as it also follows from the local dispersion relation (\ref{eq:pbm}). The behaviour of these modes in the case of negative $m$, where, as we discussed above, they are represented by a single curve with respect to $k$ (Fig. \ref{fig:buoym-1}), are similar to that for positive $m$. At $ka<1$, corresponding to the toridal mode, the growth rate increases with the absolute value of $m$ and converges to a constant value at a given $k$, while at $ka>1$, corresponding to the poloidal mode, it decreases with the absolute value of $m$, but tends to the same limiting value. 

The MRI exists only for $m=2$ and 3 in a certain interval of $k$ being most unstable at $m=2$ and $ka=0.5$. This maximum growth rate of the MRI is higher than that at $m=1$, is comparable to that of the toroidal and poloidal modes for the same $m=2$ and it does not merge with the latter as opposed to the case $m=1$ (see Fig. 11). The growth rate of the MRI at $m=3$ is decreased, its range in $k$ is narrower and shifted to larger values. The MRI disappears beyond this azimuthal wavenumber; we did not find it for larger positive $m \geq 4$ as well as for all negative $m$.    
 
\section{Summary}
\label{summary}

\begin{figure}
   \centering
   \includegraphics[width=15cm]{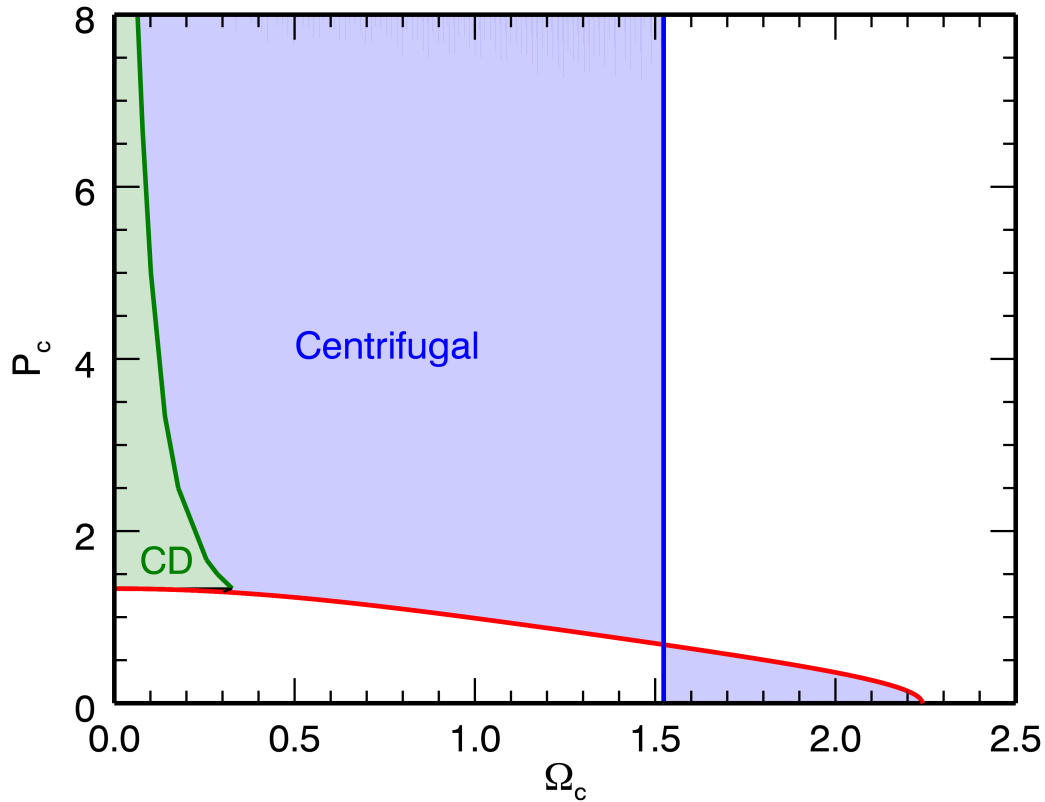} 
   \caption{\small The dominant instability types for $ka=0.1$,  in the different regions of the  parameter plane $(\Omega_c, P_c)$. In the region marked by a green shading, the instability with the largest growth rate is the CDI, while in the region marked by a blue shading the  instability with the largest growth rate is the centrifugal buoyancy instability.}
   \label{fig:summary}
\end{figure}

We have examined the stability properties of a rotating magnetized jet flow in the approximation of zero thermal pressure. Our study has focused on the effect of rotation on the CDI and on the new modes of instability introduced by rotation. In this spirit, as a first step, we did not consider the presence of a longitudinal flow, whose main effect on the modes concentrated inside the jet radius, as it is for most  of the rotationally-induced modes studied in this paper, is only that of Doppler shifting the frequency. The instability behaviour depends, of course, on the chosen equilibrium configuration and our results can be considered representative of an equilibrium configuration characterized by a distribution of current concentrated in the jet, with the return current assumed to be mainly found at very large distances. 

Similar stability analyses of rotation-induced modes in magnetized flows in the limit of zero thermal pressure are presented in \citet{Kim00, Huang03} and \citet{Pessah05}, they however make use only of a local WKB approach and the last two papers consider only axisymmetric perturbations. Our local analysis (see Section \ref{sec:wkb}) generally agrees with the results of these papers in the parameter regimes they consider. However, for our specific jet configuration  we found that the MRI in the cold limit can be present, but only in a very limited region of the parameter space and is never dominant. We extended the results of these papers to the global domain, where the WKB approach no longer holds, by solving a boundary value problem and revealed new properties of these modes that are summarized below: \\
1. At small and large axial wavenumbers $k$, the growth rates, respectively, for the toroidal and poloidal buoyancy modes obtained from the global calculations (Figs. 6, 11 and 15-18) actually exhibit a  dependence on $k$ similar to what is obtained by the corresponding local dispersion relations  (see Fig. 2). \\
2. At intermediate $k$, the behavior can be different from that predicted on the basis of the local dispersion relation, with the presence of stability gaps (for positive $m$) and merging of poloidal and toroidal modes (for negative $m$). \\
3. The properties of the MRI in the cold plasma limit, studied in the above papers based on the local dispersion relation, qualitatively agrees with our global calculations. In the nonaxisymmetric case, for positive $m$, the MRI is present only in a limited range of wavenumbers, its growth rate and the width of the unstable range reach a maximum for $ m=2$ and then (for larger $m$)  decrease (Fig. 18) consistently with \citet{Kim00},  however, the MRI is absent for $m \geq 4$ and for every negative value of $m$.  In addition,  we did not find the MRI for $m=0$ in both local and global cases, likely because of the different equilibrium adopted. In general, the MRI has always a growth rate smaller than that of the buoyancy modes.  

We have shown that two main kinds of instabilities -- CDI and buoyancy -- prevail in the considered jet flow. In Fig. \ref{fig:summary} we represent, in the parameter plane $(\Omega_c, P_c)$,  with different shadings, the regions where each of them has the largest growth rate. The figure refers to non-axisymmetric modes with $k=0.1$. We can observe that the CDI is dominant at small rotation rates and that the boundary between the CDI and centrifugal buoyancy instability regions moves towards larger values of $\Omega_c$ as we decrease $P_c$. For $P_c > 10$, the CDI  is stable for this value of the wavenumber and the only instability is the centrifugal buoyancy, which is, however, obviously stable at zero rotation. It is seen that the buoyancy instability occupies quite a large area in this parameter space in comparison with the CDI and hence should be important in jets with rotation. When we increase the wavenumber, the CDI tends to be stabilized and the centrifugal instability tends to become dominant everywhere in the parameter plane.    

Comparing now the growth rates of axisymmetric and non-axisymmetric centrifugal buoyancy modes, we see that at high wavenumbers, axisymmetric modes have a larger  growth rate, which decreases monotonically with decreasing $k$. By contrast, non-axisymmetric modes have a growth rate that is almost independent from the wavenumber and, therefore, become dominant at low values of $k$. Summarizing, at low rotation rates, the non-axisymmetric  CDI  is the instability that grows fastest and has large wavelengths. Increasing the rotation rate, the prevailing instability becomes the centrifugal axisymmetric one, which operates at small wavelengths. These results are applicable to magnetically and rotationally dominated jets, since, increasing the importance of thermal pressure, centrifugally driven modes tend to be stabilized and other modes, like pressure driven modes \citep{Kersale00} may appear. At the same time, taking into account the shear of longitudinal velocity can give rise to unstable KH modes in the jet.

This first step will be extended by introducing the effects of the longitudinal velocity also in the relativistic regime and these results will be presented in a following paper.  The different behaviour in the explored parameter space may be important  for understanding the nonlinear stages since distinct types of instability may evolve differently. This study is therefore an essential first step for the interpretation of the results of numerical simulations and for their comparison with astrophysical data.

\section*{Acknowledgments}
G.M. acknowledges the Georg Forster
Postdoctoral Research Fellowship from the Alexander von
Humboldt Foundation.

\appendix

\section{Asymptotic solution at small radii}
\label{ap:small_r}

\subsection{Case $|m| \neq 0$}
To find a solution of equations (\ref{eq:lin_system1}) and (\ref{eq:lin_system2}) at small radii, we
calculate the coefficients entering these equations at $r\rightarrow
0$ taking into account that in this limit the equilibrium quantities
$v_{0\varphi}, B_{0\varphi}\propto r$, whilst $v_{0z}$ and $B_{0z}$
tend to constant values. At $r \rightarrow 0$, these coefficients behave differently depending on whether $m=0$ or $m\neq 0$, so we should distinguish these two cases. In this subsection, we consider the case $|m|=1$ and in the next one the case $m=0$ . For $|m|=1$, we have (primes everywhere denote
radial derivative)
\[
\lim_{r\rightarrow
0}\Delta=\varrho_0(B_{0z}^2+\varrho_0c_s^2)\tilde{\omega}^4-k_B^2(B_{0z}^2+2\varrho_0c_s^2)\tilde{\omega}^2+c_s^2k_B^4,
\]
\[
\lim_{r\rightarrow
0}C_1=-\frac{2m}{r}(k_BB'_{0\varphi}+\varrho_0v'_{0\varphi}\tilde{\omega})[\tilde{\omega}^2(B_{0z}^2+\varrho_0c_s^2)-c_s^2k_B^2],
\]
\[
\lim_{r\rightarrow
0}C_2=-\frac{m^2}{r^2}[\tilde{\omega}^2(B_{0z}^2+\varrho_0c_s^2)-c_s^2k_B^2]
\]
\[
\lim_{r\rightarrow
0}C_3=\Delta(\varrho_0\tilde{\omega}^2-k_B^2)-4[\tilde{\omega}^2(\varrho_0c_s^2+B_{0z}^2)-c_s^2k_B^2](k_BB'_{0\varphi}+\varrho_0v'_{0\varphi}\tilde{\omega})^2.
\]
Substituting these coefficients into equations  (\ref{eq:lin_system1}) and (\ref{eq:lin_system2})  and
taking solutions with the form $\xi_{1r}\propto r^{\alpha},
P_1\propto r^{\alpha+1}$, to leading order, we obtain for the power
index $\alpha$,
\[
\alpha=\pm|m|-1,
\]
but because a solution must be regular at $r=0$ we choose only
$\alpha=|m|-1, (|m|\geq 1)$, and after that the ratio
\begin{equation}
\frac{P_1}{\xi_{1r}}=\frac{r}{m}[2(k_BB'_{0\varphi}+\varrho_0v'_{0\varphi}\tilde{\omega})+{\rm
sign}(m)(\varrho_0\tilde{\omega}^2-k_B^2)]
\end{equation}
This equation together with the choice $\alpha=|m|-1$ serves as our
boundary condition at small radii.

\subsection{Case $m=0$}

In the axisymmetric case $m=0$, $\Delta$ and the coefficients $C_1, C_2, C_3$  in equations (\ref{eq:lin_system1}) and (\ref{eq:lin_system2}) take the form
\[
\lim_{r\rightarrow
0}\Delta=(\varrho_0\tilde{\omega}^2-k_B^2)\left[(B_0^2+\varrho_0c_s^2)\tilde{\omega}^2-c_s^2k_B^2\right],
\]
\[
\lim_{r\rightarrow 0}C_1=r\varrho_0\tilde{\omega}^2[\tilde{\omega}^2(B^{'2}_{0\varphi}-\varrho_0v^{'2}_{0\varphi})+(\tilde{\omega}B'_{0\varphi}+v'_{0\varphi}k_B)^2],
\]
\[
\lim_{r\rightarrow 0}C_2=\varrho_0\tilde{\omega}^4-k^2[\tilde{\omega}^2(B_0^2+\varrho_0c_s^2)-c_s^2k_B^2],
\]
\[
\lim_{r\rightarrow 0}C_3=\Delta\left(\varrho_0\tilde{\omega}^2-k_B^2\right)-4[\tilde{\omega}^2(\varrho_0c_s^2+B_0^2)-c_s^2k_B^2](k_BB'_{0\varphi}+\varrho_0v'_{0\varphi}\tilde{\omega})^2,
\]
where now $\tilde{\omega}=\omega-kv_{0z}$, $k_B=kB_{0z}$ and the primes, as before, denote the radial derivatives of $v_{0\varphi}$ and $B_{0\varphi}$ at $r=0$. We see that all these coefficients are regular and finite as $r\rightarrow 0$ and only $C_1$ is proportional to $r$. 

We can express $\xi_{1r}$ through $P_1$ from equation  (\ref{eq:lin_system2})
\[
\xi_{1r}=\frac{C_1}{C_3}P_1+\frac{\Delta}{C_3}\frac{dP_1}{dr}.
\]
and substitute into equation (\ref{eq:lin_system1}). Keeping only dominant terms in the limit $r\rightarrow 0$, we get a single second order equation only for pressure
\begin{equation}
\frac{d^2P_1}{dr^2}+\frac{1}{r}\frac{dP_1}{dr}+
\frac{1}{\Delta}\left[\frac{C_1}{r}+\frac{dC_1}{dr}+\frac{C_2C_3}{\Delta}\right]P_1=0.
\end{equation}\label{pr_eq}
Since $C_1\propto r$, the coefficient in front of $P_1$ in this equation is regular and finite and explicitly calculating it at $r=0$ yields
\begin{multline*}
A\equiv\frac{1}{\Delta}\left[\frac{C_1}{r}+\frac{dC_1}{dr}+\frac{C_2C_3}{\Delta}\right]=\frac{1}{\Delta}\left[ 2\frac{dC_1}{dr}+\frac{C_2C_3}{\Delta}\right]=\frac{(\tilde{\omega}^2-c_s^2k^2)(\varrho_0\tilde{\omega}^2-k_B^2)^2}{\Delta}+\\+\frac{4B^{'2}_{0\varphi}}{\Delta}\left[\tilde{\omega}^2(\varrho_0\tilde{\omega}^2-k_B^2)+c_s^2k^2k_B^2\right] -\frac{4\varrho_0k_B\tilde{\omega}B'_{0\varphi}v'_{0\varphi}}{\Delta}\left(\tilde{\omega}^2-2c_s^2k^2\right)-\\-\frac{2\varrho_0\tilde{\omega}^2v^{'2}_{0\varphi}}{\Delta}\left( 3\varrho_0\tilde{\omega}^2-k_B^2-2\varrho_0c_s^2k^2\right),
\end{multline*}
where all the quantities in this expression are calculated at $r=0$. $A$ is nonzero constant and equation (\ref{pr_eq}) takes the form of Bessel equation of zeroth order 
\begin{equation}
\frac{d^2P_1}{dr^2}+\frac{1}{r}\frac{dP_1}{dr}+AP_1=0.
\end{equation}\label{bessel}
At $r \ll 1$, this equation has two linearly independent solutions
\[
P_1= J_0(rA^{1/2})\approx  1-\frac{A}{4}r^2, ~~~~~P_1=Y_0(rA^{1/2})\approx \frac{2}{\pi}[{\rm ln}(rA^{1/2}/2)+\gamma]J_0(rA^{1/2}),
\]
where $J_0$ and $Y_0$ are the 0-th order Bessel and Neumann functions and $\gamma=0.5772$ is the Euler-Mascheroni constant \citep{Abramowitz_Stegun72}. From these two solutions, we select the first one which is regular at small $r$: 
\[
P_1=1-\frac{A}{4}r^2
\]
and correspondingly for the displacement $\xi_{1r}$, to leading order we have 
\[
\xi_{1r}=-\frac{C_2}{2\Delta}r.
\]

\section{Asymptotic solution at large radii}
\label{ap:large_r}

To find the asymptotic behaviour of perturbations at large radii, we first derive a second order differential equation
for the pressure perturbation from equations  (\ref{eq:lin_system1}) and (\ref{eq:lin_system2})   by eliminating displacement variable $\xi_{1r}$,
\begin{equation}\label{eq:pres}
\frac{d^2P_1}{dr^2}+\left[\frac{1}{r}-\frac{\Delta}{C_3}\frac{d}{dr}\left(\frac{C_3}{\Delta}\right) \right]\frac{dP_1}{dr}+
\left[\frac{C_3}{r\Delta}\frac{d}{dr}\left(\frac{rC_1}{C_3}\right)+\frac{C_2C_3}{\Delta^2}-\frac{C_1^2}{\Delta^2}\right]P_1
=0.
\end{equation}
The equilibrium azimuthal and vertical velocities decay very quickly (exponentially) with radius, so we can put them effectively zero at asymptotically large radii, $v_{0\varphi}\approx 0, v_{0z} \approx 0$ and hence $\tilde{\omega}\approx \omega$. The equilibrium vertical magnetic field, $B_{0z}$, and density, $\varrho_0$, are constant at large radii, while the azimuthal field falls off as $B_{0\varphi}\propto 1/r$. Taking all these into account, let us estimate the coefficients entering equation (\ref{eq:pres}),
\[
\Delta=(\varrho_0\tilde{\omega}^2-k_B^2)[\tilde{\omega}^2(B_0^2+\varrho_0c_s^2)-c_s^2k_B^2],
\]
\[
C_1=2\varrho_0\tilde{\omega}^4\frac{B_{0\varphi}^2}{r}-\frac{2m}{r^2}k_BB_{0\varphi}[\tilde{\omega}^2(B_0^2+\varrho_0c_s^2)-c_s^2k_B^2] \sim O(r^{-3}),
\]
\[
C_2=\varrho_0\tilde{\omega}^4-\left(k^2+\frac{m^2}{r^2}\right)[\tilde{\omega}^2(B_0^2+\varrho_0c_s^2)-c_s^2k_B^2],
\]
\begin{multline*}
C_3=\Delta\left[\varrho_0\tilde{\omega}^2-k_B^2+r\frac{d}{dr}\left(\frac{B_{0\varphi}^2}{r^2}\right)
\right]-4k_B^2[\tilde{\omega}^2(B_0^2+\varrho_0c_s^2)-c_s^2k_B^2]\frac{B_{0\varphi}^2}{r^2}+4\varrho_0\tilde{\omega}^4\frac{B_{0\varphi}^4}{r^2}=\Delta(\varrho_0\tilde{\omega}^2-k_B^2) + O(r^{-4})
\end{multline*}
For this asymptotic expansions to be valid, the following conditions must be satisfied $r \gg H_c/B_{0z}$ and $r \gg \sqrt{H_c/kB_{0z}}$. The coefficient $C_1$ in Eqs. (\ref{eq:lin_system1}) and (\ref{eq:lin_system2})  can be assumed negligible, $C_1\approx 0$, compared with other coefficients at $r\gg (H_c/k^2B_{0z})^{1/3},~r\gg (H_c/k^3B_{0z})^{1/4}$. It easy to see that all these four inequality conditions are equivalent to two conditions: $r\gg H_c/B_{0z}$ and $r\geq 1/k$. For the derivatives we have
\[
\frac{dk_B}{dr}=\frac{m}{r}\frac{dB_{0\varphi}}{dr}-\frac{m}{r^2}B_{0\varphi}=-\frac{2m}{r^2}B_{0\varphi}\sim O(r^{-3})
\]
\[
\frac{dC_1}{dr}\simeq -6\varrho_0\tilde{\omega}^4\frac{B_{0\varphi}^2}{r^2}+\frac{6m}{r^3}k_BB_{0\varphi}[\tilde{\omega}^2(B_0^2+\varrho_0c_s^2)-c_s^2k_B^2]  \sim O(r^{-4}),
\]
\[
\frac{dC_3}{dr}\simeq -2\tilde{\omega}^2(\varrho_0\tilde{\omega}^2-k_B^2)^2\frac{B_{0\varphi}^2}{r}-2k_B[2\Delta+c_s^2(\varrho_0\tilde{\omega}^2-k_B^2)^2] \frac{dk_B}{dr} \sim O(r^{-3})
\]
Based on this, to leading order in powers of $r^{-1}$ we have
\[
\frac{\Delta}{C_3}\frac{d}{dr}\left(\frac{C_3}{\Delta}\right)\simeq-\frac{2k_B}{\varrho_0\tilde{\omega}^2-k_B^2}\frac{dk_B}{dr} \sim O(r^{-3}),
\]
\[
\frac{C_3}{r\Delta}\frac{d}{dr}\left(\frac{rC_1}{C_3} \right)=\frac{1}{r\Delta}\left(C_1+r\frac{dC_1}{dr}-\frac{rC_1}{C_3}\frac{dC_3}{dr} \right) \sim O(r^{-4}),
\]
\[
\frac{C_1^2}{\Delta^2}\sim O(r^{-6}),
\]
\[
\frac{C_2C_3}{\Delta^2}=\frac{\varrho_0\tilde{\omega}^4}{\tilde{\omega}^2(B_0^2+\varrho_0c_s^2)-c_s^2k_B^2}-k^2-\frac{m^2}{r^2}+O(r^{-4}).
\]
Thus, neglecting small terms of the order of $O(r^{-3})$ and higher in equation (\ref{eq:pres}), we obtain
\begin{equation}\label{eq:pres1}
\frac{d^2P_1}{dr^2}+\frac{1}{r}\frac{dP_1}{dr}+\left(\frac{\varrho_0\tilde{\omega}^4}{\tilde{\omega}^2(B_0^2+\varrho_0c_s^2)-c_s^2k_B^2}-k^2-\frac{m^2}{r^2}\right)P_1=0.
\end{equation}
This equation has a form similar to that of Bessel equation except the first term in brackets containing $B_{0\varphi}$ (through $B_0^2$ and $k_B$) and therefore depending on $r$. However, in our shooting method, to perform a backward intergation from large to smaller radii, we need an analytical solution of equation (\ref{eq:pres1}). To this end,  at very large radii, in the transition region $r_1\leq r \leq r_2$, where $r_1=30$ and $r_2=50$, we impose a return current, which at $r>r_2$ cancels the azimuthal field and maintains constant $B_{0z}$ (this procedure does not affect the eigenfunctions and growth rates of the unstable modes). In this case, at these radii, equation (\ref{eq:pres1}) exactly matches the Bessel differential equation
\[
\frac{d^2P_1}{dr^2}+\frac{1}{r}\frac{dP_1}{dr}+\left(\chi^2-\frac{m^2}{r^2} \right)P_1=0,
\]
where the parameter
\[
\chi=\sqrt{\frac{\varrho_0\tilde{\omega}^4}{\tilde{\omega}^2(B_0^2+\varrho_0c_s^2)-c_s^2k_B^2}-k^2}
\]
is generally complex and does not depend on radius. Solutions to this equation can be represented as the Hankel functions of the order $m$ with the following asymptotic forms at $r\rightarrow \infty$ \citep{Abramowitz_Stegun72},
\[
P_1=H_{m}^{(1)}(\chi r)  \sim \sqrt{\frac{2}{\pi \chi r}}\exp\left[{\rm i} \left(\chi r-\frac{m\pi}{2}-\frac{\pi}{4} \right) \right],~~~~P_1=H_{m}^{(2)}(\chi r)  \sim \sqrt{\frac{2}{\pi \chi r}}\exp\left[-{\rm i} \left(\chi r-\frac{m\pi}{2}-\frac{\pi}{4} \right) \right].
\]
Depending on the sign of the imaginary part of $\chi$ one of these two solutions is selected: if $Im(\chi)>0$ then $P_1=H_{m}^{(1)}(\chi r)$, whereas if $Im(\chi)<0$ then $P_1=H_{m}^{(2)}(\chi r)$, so that in both cases the pressure perturbation exponentially decays with radis. In addition, we also require that the solution at large radii correspond to outgoing waves (Sommerfeld condition), which implies that an eigen-$\omega$ must satisfy  $Re(\omega)\cdot Re(\chi)<0$. 

With the above asymptotic form of $P_1$, one can readily find the displacement
\[
\xi_{1r}=\frac{\Delta}{C_3}\frac{dP_1}{dr}+\frac{C_1}{C_3}P_1\simeq \frac{1}{\varrho_0\tilde{\omega}^2-k_B^2}\frac{dP_1}{dr}+O(r^{-3})=\frac{P_1}{\varrho_0\tilde{\omega}^2-k_B^2}\left(\pm {\rm i}\chi-\frac{1}{2r}\right)+O(r^{-3}).
\]

\label{lastpage}


\begin{thebibliography}{}

\bibitem[\protect\citeauthoryear{{Abramowitz} \& {Stegun}}{{Abramowitz} \&
  {Stegun}}{1972}]{Abramowitz_Stegun72}
{Abramowitz} M.,  {Stegun} I.~A.,  1972, {Handbook of Mathematical Functions,
  New York: Dover}

\bibitem[\protect\citeauthoryear{{Appl}}{{Appl}}{1996}]{Appl96}
{Appl} S.,  1996, \aap, 314, 995

\bibitem[\protect\citeauthoryear{{Appl} \& {Camenzind}}{{Appl} \&
  {Camenzind}}{1992}]{Appl92}
{Appl} S.,  {Camenzind} M.,  1992, \aap, 256, 354

\bibitem[\protect\citeauthoryear{{Appl}, {Lery} \& {Baty}}{{Appl}
  et~al.}{2000}]{Appl00}
{Appl} S.,  {Lery} T.,    {Baty} H.,  2000, \aap, 355, 818

\bibitem[\protect\citeauthoryear{{Balbus} \& {Hawley}}{{Balbus} \&
  {Hawley}}{1992}]{Balbus92}
{Balbus} S.~A.,  {Hawley} J.~F.,  1992, \apj, 400, 610

\bibitem[\protect\citeauthoryear{{Baty} \& {Keppens}}{{Baty} \&
  {Keppens}}{2002}]{Baty02}
{Baty} H.,  {Keppens} R.,  2002, \apj, 580, 800

\bibitem[\protect\citeauthoryear{{Begelman}}{{Begelman}}{1998}]{Begelman98}
{Begelman} M.~C.,  1998, \apj, 493, 291

\bibitem[\protect\citeauthoryear{{Birkinshaw}}{{Birkinshaw}}{1991}]{Birkinshaw91}
{Birkinshaw} M.,  1991, {The stability of jets}.
p.~278

\bibitem[\protect\citeauthoryear{{Blokland}, {van der Swaluw}, {Keppens} \&
  {Goedbloed}}{{Blokland} et~al.}{2005}]{Blokland05}
{Blokland} J.~W.~S.,  {van der Swaluw} E.,  {Keppens} R.,    {Goedbloed} J.~P.,
   2005, \aap, 444, 337

\bibitem[\protect\citeauthoryear{{Bodo}, {Mamatsashvili}, {Rossi} \&
  {Mignone}}{{Bodo} et~al.}{2013}]{Bodo13}
{Bodo} G.,  {Mamatsashvili} G.,  {Rossi} P.,    {Mignone} A.,  2013, \mnras,
  434, 3030

\bibitem[\protect\citeauthoryear{{Bodo}, {Rosner}, {Ferrari} \&
  {Knobloch}}{{Bodo} et~al.}{1989}]{Bodo89}
{Bodo} G.,  {Rosner} R.,  {Ferrari} A.,    {Knobloch} E.,  1989, \apj, 341, 631

\bibitem[\protect\citeauthoryear{{Bodo}, {Rosner}, {Ferrari} \&
  {Knobloch}}{{Bodo} et~al.}{1996}]{Bodo96}
{Bodo} G.,  {Rosner} R.,  {Ferrari} A.,    {Knobloch} E.,  1996, \apj, 470, 797

\bibitem[\protect\citeauthoryear{{Bonanno} \& {Urpin}}{{Bonanno} \&
  {Urpin}}{2006}]{Bonanno06}
{Bonanno} A.,  {Urpin} V.,  2006, \pre, 73, 066301

\bibitem[\protect\citeauthoryear{{Bonanno} \& {Urpin}}{{Bonanno} \&
  {Urpin}}{2007}]{Bonanno07}
{Bonanno} A.,  {Urpin} V.,  2007, \apj, 662, 851

\bibitem[\protect\citeauthoryear{{Bonanno} \& {Urpin}}{{Bonanno} \&
  {Urpin}}{2008}]{Bonanno08}
{Bonanno} A.,  {Urpin} V.,  2008, \aap, 488, 1

\bibitem[\protect\citeauthoryear{{Bonanno} \& {Urpin}}{{Bonanno} \&
  {Urpin}}{2011a}]{Bonanno11}
{Bonanno} A.,  {Urpin} V.,  2011a, \pre, 84, 056310

\bibitem[\protect\citeauthoryear{{Bonanno} \& {Urpin}}{{Bonanno} \&
  {Urpin}}{2011b}]{Bonanno11a}
{Bonanno} A.,  {Urpin} V.,  2011b, \aap, 525, A100

\bibitem[\protect\citeauthoryear{{Bondeson}, {Iacono} \&
  {Bhattacharjee}}{{Bondeson} et~al.}{1987}]{Bondeson87}
{Bondeson} A.,  {Iacono} R.,    {Bhattacharjee} A.,  1987, Physics of Fluids,
  30, 2167

\bibitem[\protect\citeauthoryear{{Carey} \& {Sovinec}}{{Carey} \&
  {Sovinec}}{2009}]{Carey09}
{Carey} C.~S.,  {Sovinec} C.~R.,  2009, \apj, 699, 362

\bibitem[\protect\citeauthoryear{{Ferrari}, {Trussoni} \&
  {Zaninetti}}{{Ferrari} et~al.}{1978}]{Ferrari78}
{Ferrari} A.,  {Trussoni} E.,    {Zaninetti} L.,  1978, \aap, 64, 43

\bibitem[\protect\citeauthoryear{{Freidberg}}{{Freidberg}}{1987}]{Freidberg87}
{Freidberg} J.,  1987, {Ideal Magnetohydrodynamics, Plenum Press, New York}

\bibitem[\protect\citeauthoryear{{Frieman} \& {Rotenberg}}{{Frieman} \&
  {Rotenberg}}{1960}]{Frieman60}
{Frieman} E.,  {Rotenberg} M.,  1960, Reviews of Modern Physics, 32, 898

\bibitem[\protect\citeauthoryear{{Fu} \& {Lai}}{{Fu} \& {Lai}}{2011}]{Fu11}
{Fu} W.,  {Lai} D.,  2011, \mnras, 410, 399

\bibitem[\protect\citeauthoryear{{Goedbloed}}{{Goedbloed}}{2009}]{Goedbloed09}
{Goedbloed} J.~P.,  2009, Physics of Plasmas, 16, 122110

\bibitem[\protect\citeauthoryear{Goedbloed, Keppens \& Poedts}{Goedbloed
  et~al.}{2010}]{Goedbloed10}
Goedbloed J.~P.,  Keppens R.,    Poedts S.,  2010, Advanced
  magnetohydrodynamics : with applications to laboratory and astrophysical
  plasmas.
Cambridge University Press, Cambridge; New York

\bibitem[\protect\citeauthoryear{{Hanasz}, {Sol} \& {Sauty}}{{Hanasz}
  et~al.}{2000}]{Hanasz00}
{Hanasz} M.,  {Sol} H.,    {Sauty} C.,  2000, \mnras, 316, 494

\bibitem[\protect\citeauthoryear{{Hardee}}{{Hardee}}{1979}]{Hardee79}
{Hardee} P.~E.,  1979, \apj, 234, 47

\bibitem[\protect\citeauthoryear{{Hardee}}{{Hardee}}{2006}]{Hardee06}
{Hardee} P.~E.,  2006, in {P.~A.~Hughes \& J.~N.~Bregman} ed., Relativistic
  Jets: The Common Physics of AGN, Microquasars, and Gamma-Ray Bursts Vol.~856
  of American Institute of Physics Conference Series, {AGN Jets: A Review of
  Stability and Structure}.
pp 57--77

\bibitem[\protect\citeauthoryear{{Hardee}, {Cooper}, {Norman} \&
  {Stone}}{{Hardee} et~al.}{1992}]{Hardee92}
{Hardee} P.~E.,  {Cooper} M.~A.,  {Norman} M.~L.,    {Stone} J.~M.,  1992,
  \apj, 399, 478

\bibitem[\protect\citeauthoryear{{Huang} \& {Hassam}}{{Huang} \&
  {Hassam}}{2003}]{Huang03}
{Huang} Y.-M.,  {Hassam} A.~B.,  2003, Physics of Plasmas, 10, 204

\bibitem[\protect\citeauthoryear{{Istomin} \& {Pariev}}{{Istomin} \&
  {Pariev}}{1994}]{Pariev94}
{Istomin} Y.~N.,  {Pariev} V.~I.,  1994, \mnras, 267, 629

\bibitem[\protect\citeauthoryear{{Istomin} \& {Pariev}}{{Istomin} \&
  {Pariev}}{1996}]{Pariev96}
{Istomin} Y.~N.,  {Pariev} V.~I.,  1996, \mnras, 281, 1

\bibitem[\protect\citeauthoryear{{Keppens}, {Casse} \& {Goedbloed}}{{Keppens}
  et~al.}{2002}]{Keppens02}
{Keppens} R.,  {Casse} F.,    {Goedbloed} J.~P.,  2002, \apjl, 569, L121

\bibitem[\protect\citeauthoryear{{Kersal{\'e}}, {Longaretti} \&
  {Pelletier}}{{Kersal{\'e}} et~al.}{2000}]{Kersale00}
{Kersal{\'e}} E.,  {Longaretti} P.-Y.,    {Pelletier} G.,  2000, \aap, 363,
  1166

\bibitem[\protect\citeauthoryear{{Kim}, {Balsara}, {Lyutikov}, {Komissarov},
  {George} \& {Siddireddy}}{{Kim} et~al.}{2015}]{Kim15}
{Kim} J.,  {Balsara} D.~S.,  {Lyutikov} M.,  {Komissarov} S.~S.,  {George} D.,
    {Siddireddy} P.~K.,  2015, \mnras, 450, 982

\bibitem[\protect\citeauthoryear{{Kim} \& {Ostriker}}{{Kim} \&
  {Ostriker}}{2000}]{Kim00}
{Kim} W.-T.,  {Ostriker} E.~C.,  2000, \apj, 540, 372

\bibitem[\protect\citeauthoryear{{Lyubarskii}}{{Lyubarskii}}{1999}]{Lyubarski99}
{Lyubarskii} Y.~E.,  1999, \mnras, 308, 1006

\bibitem[\protect\citeauthoryear{{Mizuno}, {Hardee} \& {Nishikawa}}{{Mizuno}
  et~al.}{2007}]{Hardee07}
{Mizuno} Y.,  {Hardee} P.,    {Nishikawa} K.-I.,  2007, \apj, 662, 835

\bibitem[\protect\citeauthoryear{{Narayan}, {Li} \& {Tchekhovskoy}}{{Narayan}
  et~al.}{2009}]{Narayan09}
{Narayan} R.,  {Li} J.,    {Tchekhovskoy} A.,  2009, \apj, 697, 1681

\bibitem[\protect\citeauthoryear{{Perucho}, {Hanasz}, {Mart{\'{\i}}} \&
  {Sol}}{{Perucho} et~al.}{2004}]{Perucho04}
{Perucho} M.,  {Hanasz} M.,  {Mart{\'{\i}}} J.~M.,    {Sol} H.,  2004, \aap,
  427, 415

\bibitem[\protect\citeauthoryear{{Perucho}, {Mart{\'{\i}}}, {Cela}, {Hanasz},
  {de La Cruz} \& {Rubio}}{{Perucho} et~al.}{2010}]{Perucho10}
{Perucho} M.,  {Mart{\'{\i}}} J.~M.,  {Cela} J.~M.,  {Hanasz} M.,  {de La Cruz}
  R.,    {Rubio} F.,  2010, \aap, 519, A41+

\bibitem[\protect\citeauthoryear{{Pessah} \& {Psaltis}}{{Pessah} \&
  {Psaltis}}{2005}]{Pessah05}
{Pessah} M.~E.,  {Psaltis} D.,  2005, \apj, 628, 879

\bibitem[\protect\citeauthoryear{{Sikora}, {Begelman}, {Madejski} \&
  {Lasota}}{{Sikora} et~al.}{2005}]{Sikora05}
{Sikora} M.,  {Begelman} M.~C.,  {Madejski} G.~M.,    {Lasota} J.-P.,  2005,
  \apj, 625, 72

\bibitem[\protect\citeauthoryear{{Tomimatsu}, {Matsuoka} \&
  {Takahashi}}{{Tomimatsu} et~al.}{2001}]{Tomimatsu01}
{Tomimatsu} A.,  {Matsuoka} T.,    {Takahashi} M.,  2001, \prd, 64, 123003

\bibitem[\protect\citeauthoryear{{Urpin}}{{Urpin}}{2002}]{Urpin02}
{Urpin} V.,  2002, \aap, 385, 14

\bibitem[\protect\citeauthoryear{{Varni{\`e}re} \& {Tagger}}{{Varni{\`e}re} \&
  {Tagger}}{2002}]{Varniere02}
{Varni{\`e}re} P.,  {Tagger} M.,  2002, \aap, 394, 329

\end{thebibliography}
\end{document}